\documentclass[twocolumn,showpacs,aps,pra,amsmath,amssymb]{revtex4-2}

\usepackage{mathrsfs}
\usepackage{amsfonts}
\usepackage{amsmath,amsfonts,amssymb,mathrsfs,bm}
\usepackage{verbatim,psfrag,times,CJK}
\usepackage{graphicx,graphics,color,epsfig}
\usepackage{float}
\DeclareMathOperator\arctanh{arctanh}
\usepackage{flafter}
\usepackage{subeqnarray}
\usepackage{cases}
\usepackage{amsmath, upgreek, amssymb, graphicx, subfigure, paralist, dsfont}
\usepackage[colorlinks, linkcolor=blue, citecolor=blue, urlcolor=blue, breaklinks=true]{hyperref}

\definecolor{red}{rgb}{1,0,0}

\begin{document}

\title{Effect of linear and nonlinear coupling processes on correlation properties of bosonic modes}

\author{Jakub Bembenek}
\author{Zbigniew Ficek}
\email{z.ficek@if.uz.zgora.pl}

\address{Quantum Optics and Engineering Division, Institute of Physics, University of Zielona G\'ora, Szafrana 4a, 65-516 Zielona G\'ora, Poland}

\date{\today}

%\corres{Correspondence: z.ficek@if.uz.zgora.pl}

%\doi{\url{}}

\begin{abstract}
We study the influence of the linear and nonlinear coupling processes on the correlation properties of a system composed of two bosonic modes to highlight the similarities and differences between these two types of coupling. The coupling processes are treated as Gaussian while the modes are assumed to experience damping and fluctuations that are due to their coupling to independent squeezed vacua. Analytic expressions are obtained for the variances and populations of the modes, single-mode two-photon correlation functions, and the inter-mode one-photon correlation function which carries  information about coherence properties of the modes, and the two-photon correlation function which reveals a quantum feature in that the correlation is necessary for entanglement between the modes. We investigate under what circumstances a given type of coupling causes the creation of the first-order correlations and which causes the creation of the two-photon correlations. Distinctly different results are obtained for the linear and nonlinear couplings, especially when there are two-photon correlations present in the modes. We identify the connection between variation of the single mode populations and correlations, and the creation of the two-mode correlations. In particular, the linear coupling process can generate the inter-mode correlations only when there are differences between the modes either in populations or two-photon correlations. The varying with the coupling strength a population difference is shown to be responsible for generation of the first-order correlations while the varying asymmetry in the two-photon correlations is found to be responsible for the generation of the inter-mode two-photon correlations. In the case of the nonlinear coupling the generation of the inter-mode correlations is insensitive to any difference between the modes. The varying with the coupling strength amplification of the population of the modes is found to be responsible for generation of the inter-mode two photon correlations while the varying amplification of the two-photon correlations results in the first-order correlations between the modes. Furthermore, in the strong coupling limit the linear coupling tends to destroy all of the inter-mode correlations, and simultaneously turns the states of the modes to be identical, either thermal or equally squeezed states. In the case of nonlinear coupling process the modes are turned to be perfectly coherent with all the two-photon correlations reduced to the limit of maximal classical correlations. 
\end{abstract}

\maketitle

\section{Introduction}\label{sec1}

There have been a significant interest in implementation of quantum networks consisting of a set of spatially separated nodes, e.g., single mode cavities, connected by a coupling process appearing as a quantum communication channel~\cite{cz97,np24,hd24}. Operation of such systems demands a careful design of the coupling process between the nodes and preparation of the nodes in suitable states. The coupling process may create coherence and/or multi-photon correlations between nodes which are crucial for the transfer of an information between the nodes. 

The problem of creation of coherence and multi-photon correlations between quantum systems has attracted considerable interest over the years not only because of a basic desire to understand how coherence and correlations could be created but also because of their importance in determination of nonclassical states of quantum systems~\cite{rjg63,rg63,mw95,fs04,sl12,lg16}. Various types of correlations can exist between quantum systems and their importance in understanding properties of quantum systems is often discussed in connection with different phenomena. For example, interference and quantum beats are amongst the simplest examples of phenomena resulting from the presence of mutual coherence, the so-called first-order correlation between quantum systems. Nonclassical phenomena such as squeezing and entanglement result from presence of a different kind of correlations, often referred to as anomalous correlations~\cite{ga86,hr87,af24}. 
The mutual coherence resulting from the first-order correlation is produced by a constant or nearly constant phase difference between quantum systems~\cite{mw95,fs04,ww16,ol19}. There are, however, coherence effects resulting from higher-order correlations, e.g. the intensity correlations, which are possible even when the phase difference between systems is random~\cite{lm83,gh86,sb14,pf15}.

In this paper we examine correlation properties of the basic element of a quantum network, a pair of bosonic modes, each coupled to own reservoir, and subjected to the mutual interaction by a coupling process. Two types of the coupling processes are considered, the linear coupling process based on exchange of single photons or on mixing modes at a beamsplitter, and the nonlinear two-photon coupling process based on parametric down conversion. We are particularly interested in the manner the coupling processes create inter-mode correlation and how the coupling affects initial correlation properties of the modes. We show that the linear coupling process can create correlations among the modes only if they are at least partly distinguishable in either the number of photons or degree of the two-photon correlations or mutual orientation of the noise ellipses. If this condition is not fulfilled the process does not couple the modes. In contrast, the nonlinear process is less restrictive to the state of the single modes and couples the modes when they are indistinguishable. We find that there are significant differences in the results predicted by the two coupling processes. In particular, the presence of the two-photon correlations inside the modes has nothing to do with the creation of the two-photon inter-modes correlations. The creation of the correlations is intimately connected to the amplification of the population of the modes. We also identify the connection between the creation of the inter-mode correlations and the variation of the population of the modes and the single-mode two-photon correlations.

The correlation properties of a two-mode system under linear and nonlinear processes have not, to out knowledge, been investigated in such details. In particular, the emphasis is upon the comparison of the correlation properties of the system under the linear and nonlinear interactions and to determine conditions for the maximal inter-mode correlations.  It may be noted that it is known for years that the nonlinear parametric down conversion process results in an entanglement of the two output modes~\cite{bk85,wk86,ow89,rk94}. Creation of entanglement using a beamsplitter has also been studied~\cite{bs92,yb93,pa99,sw01,ks02,hf06,ti09,bb15,gt15}. However, these calculations have been specifically oriented towards studying the generation of entanglement between the modes. As such, they miss the problem of the creation of the first-order coherence, which is instrumental in continuous variable quantum computation. In addition, they miss the problem of the influence of the coupling processes on single-mode states, which can be significantly modified by the coupling processes. 

The paper is organized as follows. In Sec.~\ref{sec2} we formulate our model, introduce correlation functions and degrees of correlations whose properties under the presence of a coupling between the modes will be studied in details. Our major purpose is to compare the correlation properties of the system for two physically different types of coupling processes. In Sec.~\ref{sec3}, we specialize to the linear coupling process and the dependence of the correlation properties on the strength of the coupling is discussed for various states the modes are prepared prior the mutual coupling.  We pay particular attention to conditions under which the coupling leads to the creation of the inter-mode one- and two-photon correlations. The effect of the interaction on the populations of the modes and the single-mode two-photon correlations is also investigated. Then, in Sec.~\ref{sec4}, we examine correlation properties of the system under the nonlinear coupling process. In Sec.~\ref{sec5}, we determine relationship between the creation of inter-mode correlations and the variation of the single-mode populations and correlations. A summary of the results is contained in Sec.~\ref{sec6}.
A detailed solution of the Heisenberg-Langevin equations and the outline of the method used in the evaluation of the correlation functions are presented in the Appendix.

\section{Model}\label{sec2}

We consider a pair of mutually coupled bosonic modes each interacting with its surrounding environment. The modes are described by bosonic creation (annihilation) operators, $a^{\dag} (a)$ and $b^{\dag} (b)$,  respectively. The bosonic operators associated with the same mode satisfy the usual commutation relation $[r,r^{\dag}]=1,\, (r=a,b)$ and at the initial time $t=0$ the operators of different modes commute. We assume that the modes are coupled to each other and concentrate on two types of the coupling processes, the linear process of the beamsplitter or photon exchange type interaction, which is determined by the interaction Hamiltonian $(\hbar =1)$
\begin{align}
H_{L} = -ig\left(a^{\dag}be^{-i(\phi_{a}-\phi_{b})} - a b^{\dag}e^{i(\phi_{a}-\phi_{b})}\right)  ,\label{e1}
\end{align}
and the nonlinear process of the parametric down conversion type determined by the Hamiltonian
\begin{align}
H_{NL} = -ig\left(abe^{i(\phi_{a}+\phi_{b})} - a^\dag b^{\dag}e^{-i(\phi_{a}+\phi_{b})}\right)  ,\label{e2}
\end{align}
where $g$, assumed to be real, determines the strength of the coupling between the modes, and $\phi_{i}\, (i=a,b)$ is the phase of the $i$-the mode. The action of the Hamiltonians will couple the modes and will give the systematic evolution of the modes due to the mutual interaction. Without loss of generality we will assume that the phase of the mode $b$ is fixed at $\phi_{b}=0$, while the phase $\phi_{a}\equiv \phi$ of the mode $a$ can be varied.

There are several examples of practical optical and microwave systems in which the pure linear and nonlinear mode coupling interactions can be implemented between two bosonic modes. For example, the linear coupling can be implemented by sending two modes into a beamsplitter with the mode and beamsplitter parameters chosen to be consistent with the Hamiltonian (\ref{e1}), see Ref.~\cite{bsr92} for a practical implementation. Another example, polariton modes of two spatially separated microcavity quantum boxes being close enough together such that
particles can tunnel from one box to the other~\cite{ls10}. An alternative way is to propagate fields in linearly coupled waveguides or linearly coupled optical cavities, or in single-mode optomechanical cavities where the cavity and mechanical modes serve as two linearly coupled modes. 
According to the nonlinear process described by the Hamiltonian (\ref{e2}), there are several practical examples of how to create nonlinear coupling between two systems~\cite{mm19,ak13,ak14,ps06,tx09,lp16}. For example, Menotti {\it et al.}~\cite{mm19} have experimentally realised coupling solely through a nonlinear interaction between two racetrack resonators. The pure nonlinear interaction can be realized also in several different ways in superconducting quantum circuits~\cite{ak13}. Optomechanical systems can also be prepared to form a nonlinear radiation-pressure coupling between the mechanical and cavity modes and thus present another possible realization of the nonlinear coupling~\cite{ak14}.

We are interested, primarily, in the variation of the fluctuations, populations and the two-photon correlations of two Gaussian modes under the applied interaction, and in the identification of conditions for the creation of the one- and two-photon correlations between the modes. The fluctuations and the correlations not only allow to determine statistical properties of the fields but also allow one to discriminate between quantum and classical fields. Properties of the modes are determined by their populations, $\langle a^{\dag}a\rangle$ and $\langle b^{\dag}b\rangle$, two-photon phase dependent correlations, $\langle aa\rangle$ and $\langle bb\rangle$, and fluctuations described by the variances $\langle \Delta^{2}X_{i}\rangle =\langle X_{i}^{2}\rangle$ and $\langle \Delta^{2}Y_{i}\rangle = \langle Y_{i}^{2}\rangle$ of the real in-phase $X_{i}$ and out-off-phase $Y_{i}\, (i=a,b)$ quadrature components of the bosonic operators, $a =(X_{a}+iY_{a})e^{-i\phi}/\sqrt{2}, b=(X_{b}+iY_{b})/\sqrt{2}$. The mutual correlation properties of the Gaussian fields are described by the cross-correlation functions, the first-order correlation function $\langle a^{\dag}b\rangle$ and the two-photon cross-correlation function $\langle ab\rangle$. The fact that $\langle a^{\dag}b\rangle\neq 0$ means that the modes are at least partially coherent whereas $\langle ab\rangle\neq 0$ is necessary but not sufficient for the modes to be entangled.

To quantify the degree of coherence and to discriminate between quantum and classical fields we will consider properties of the normalized single-mode correlation functions
\begin{align}
\eta_{aa} &= \frac{|\langle aa\rangle|}{\langle a^{\dag}a\rangle} ,\quad \eta_{bb} = \frac{|\langle bb\rangle|}{\langle b^{\dag}b\rangle} ,\label{e3}
\end{align}
and two-mode correlation functions
\begin{align}
\gamma_{ab} &= \frac{|\langle a^{\dag}b\rangle|}{\sqrt{\langle a^{\dag}a\rangle\langle b^{\dag}b\rangle}} ,\quad \eta_{ab} = \frac{|\langle ab\rangle|}{\sqrt{\langle a^{\dag}a\rangle\langle b^{\dag}b\rangle}} .\label{e4}
\end{align}
Nonzero values of $\eta_{ii}\, (i=a,b)$ functions results in asymmetric distribution of fluctuations (noise) between quadrature components of the fields that the fluctuations of one of the quadrature components are enhanced in expense of the fluctuations of the other component which exhibits squeezed fluctuations. There is a threshold value of $\eta_{ii}=1$ which discriminates between classically and quantum squeezed fluctuations of the $i$-the field that values of $\eta_{ii}<1$ indicate the classically squeezed field since in this case the noise level of the squeezed quadrature of the field is above the shot noise level, while values of $\eta_{ii}>1$ indicate the quantum squeezed field since in this case the noise level of the squeezed quadrature falls below the shot noise level. 

The correlation function $\gamma_{ab}\in (0,1)$ that $\gamma_{ab}=0$ means the perfectly incoherent while $\gamma_{ab}=1$ means perfectly coherent modes. The correlation function $\eta_{ab}\in (0,\infty)$ but there is a threshold value of $\eta_{ab}$ at which one can distinguish between the cases of classically and quantum correlated (entangled) modes. In the literature there are several different criteria for entanglement applicable for Gaussian systems~\cite{hh96,dg00,rs00,hg03,sw05,hz06}. For example, the DGCZ~\cite{dg00} and Hillery-Zubairy~\cite{hz06} criteria show that the necessary and sufficient condition for entanglement of two vacuum or thermal modes is $\eta_{ab}>1$. The criterion differs from those two mentioned when the modes contain two-photon correlations $(\eta_{aa}=\eta_{bb}\neq 0)$ and/or are partly coherent $(\gamma_{ab}\neq 0)$.
Referring to the criterion of a quantum field involving the violation of the Cauchy-Schwartz inequality, it is not difficult to show that for two Gaussian modes containing two-photon correlations and/or being partly coherent $(\gamma_{ab}\neq 0)$, the criterion for entanglement of the modes~is
\begin{align}
\eta_{ab}^{2} > 1+\eta_{aa}^{2}-\gamma_{ab}^{2} .\label{e5}
\end{align}
The degree of entanglement will not concern us but it might be determined evaluating the eigenvalues of the covariance matrix.

In what follows we evaluate single and two-mode correlation functions and fluctuations of the modes for the two types of the interaction between the modes described by the interaction Hamiltonians (\ref{e1}) and (\ref{e2}), and discuss in details their properties for three sets of states the modes are prepared prior to the mutual interaction. In the first, both input modes are in equally squeezed states. In the second, the modes are in equally populated but their fluctuations unequally squeezed, and in the third the field of one of the modes is squeezed while the field of the other is in vacuum. In the case of equally squeezed fields of the input modes we assume that the noise ellipse of one of the mode, mode $a$ can be rotated with an angle $\phi$ relative to the orientation of the noise ellipse of the other mode.

\begin{figure}[h]
  \includegraphics[width=0.45\textwidth]{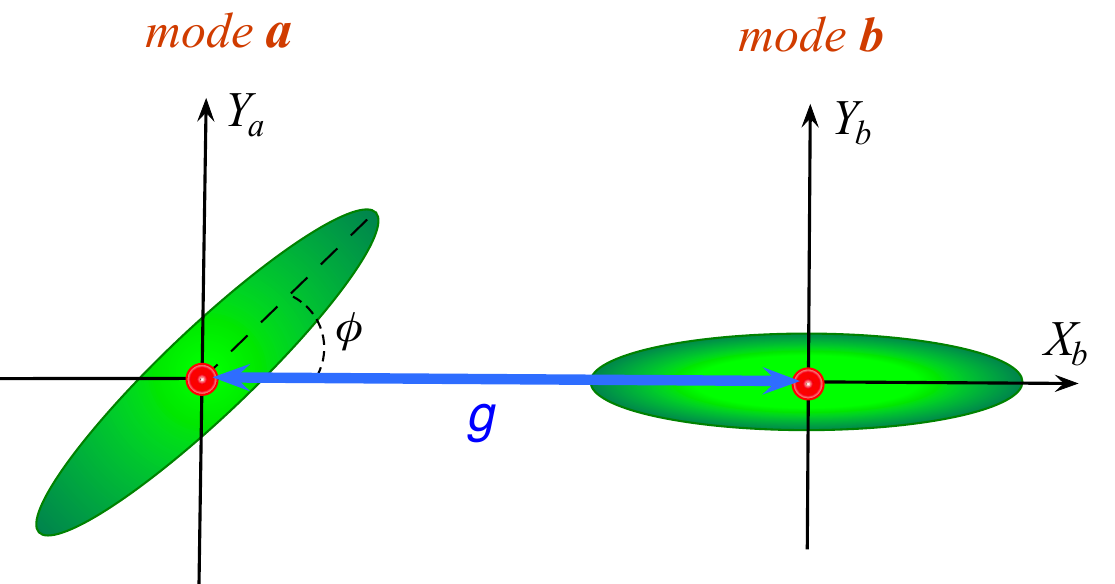}
  \caption{Illustration od noise ellipses of two independently squeezed modes which can mutually interact with a coupling strength $g$. The noise ellipse of mode $a$ is allowed to rotate with the relative phase $\phi$.}
  \label{fig1}
\end{figure}

Employing the input-output formalism~\cite{gc85,by04,gz04} it is straightforward to show that the quadrature components obey the following set of coupled Heisenberg-Langevin equations
\begin{align}
\dot{X}_{a} &= -i\left[X_{a},H_{i}\right] -\kappa X_{a} -\sqrt{2\kappa}X_{a}^{{\rm in}}(t) ,\nonumber\\
\dot{Y}_{a} &= -i\left[Y_{a},H_{i}\right] -\kappa Y_{a} -\sqrt{2\kappa}Y_{a}^{{\rm in}}(t) ,\nonumber\\
\dot{X}_{b} &= -i\left[X_{b},H_{i}\right] -\kappa X_{b} -\sqrt{2\kappa}X_{b}^{{\rm in}}(t) ,\nonumber\\
\dot{Y}_{b} &= -i\left[Y_{b},H_{i}\right] -\kappa Y_{b} -\sqrt{2\kappa}Y_{b}^{{\rm in}}(t) ,\label{e6}
\end{align}
where $H_{i}$ is either linear or nonlinear interaction Hamiltonian and $\kappa$ is the photon loss rate of the modes, assumed the same for both modes. $X_{i}^{{\rm in}}(t), Y_{j}^{{\rm in}}(t)\, (i,j=a,b)$ are Langevin noise terms representing spontaneous emission fluctuations. They are taken to be $\delta$ correlated and Gaussian, with
\begin{align}
\langle X_{a}^{{\rm in}}(t)X_{a}^{{\rm in}}(t^{\prime})\rangle &= \left(\frac{1}{2}+n_{a}+m_{a}\cos2\phi\right)\delta(t-t^{\prime}) ,\nonumber \\
\langle Y_{a}^{{\rm in}}(t)Y_{a}^{{\rm in}}(t^{\prime})\rangle &= \left(\frac{1}{2}+n_{a}-m_{a}\cos2\phi\right)\delta(t-t^{\prime}) ,\nonumber \\
\langle X_{a}^{{\rm in}}(t)Y_{a}^{{\rm in}}(t^{\prime})\rangle &= -\langle Y_{a}^{{\rm in}}(t)X_{a}^{{\rm in}}(t^{\prime})\rangle \nonumber\\
&= \left(\frac{1}{2}i-m_{a}\sin2\phi\right)\delta(t-t^{\prime}) ,\nonumber \\
\langle X_{b}^{{\rm in}}(t)X_{b}^{{\rm in}}(t^{\prime})\rangle &= \left(\frac{1}{2}+n_{b}+m_{b}\right)\delta(t-t^{\prime}) ,\nonumber \\
\langle Y_{b}^{{\rm in}}(t)Y_{b}^{{\rm in}}(t^{\prime})\rangle &= \left(\frac{1}{2}+n_{b}-m_{b}\right)\delta(t-t^{\prime}) ,\nonumber\\
\langle X_{b}^{{\rm in}}(t)Y_{b}^{{\rm in}}(t^{\prime})\rangle &= -\langle Y_{b}^{{\rm in}}(t)X_{b}^{{\rm in}}(t^{\prime})\rangle = \frac{1}{2}i\delta(t-t^{\prime}),\label{e7}
\end{align}
and 
\begin{align}
&\langle X_{i}^{{\rm in}}(t)X_{j}^{{\rm in}}(t^{\prime})\rangle = \langle Y_{i}^{{\rm in}}(t)Y_{j}^{{\rm in}}(t^{\prime})\rangle \nonumber \\
&= \langle X_{i}^{{\rm in}}(t)Y_{j}^{{\rm in}}(t^{\prime})\rangle = \langle Y_{i}^{{\rm in}}(t)X_{j}^{{\rm in}}(t^{\prime})\rangle = 0 , \, i\neq j  ,\label{e7a}
\end{align}
where $n_{i}$ is the number of thermal photons present in the $i$-the mode, $m_{i}$ is the degree of two-photon correlations between photons of the $i$-th mode, $\phi$ is the relative phase between the noise ellipses of the modes. We assume that directions of the principal axis of the noise ellipse of the mode $a$ can be rotated with $\phi$, whereas the directions of the axis of the noise ellipse of the mode $b$ are fixed, as illustrated in Fig.~\ref{fig1}. The parameters $n_{i}$ and $m_{i}$ are related to each other such that $m_{i}\leq n_{i}$ corresponds to a classically squeezed field whereas $m_{i}\leq \sqrt{n_{i}(n_{i}+1)}$ corresponds to a quantum squeezed field. It is reflected in the values of the correlation functions, either $X_{i}$ or $Y_{i}$ quadratures can be reduced below one-half when a given mode is in quantum squeezed state. It is easy to see that this can be achieved when $\eta_{ii}=m_{i}/n_{i}>1$.

In the following sections, the solution of Eqs.~(\ref{e6}) will be applied to determine variation of the variances of the modes and the intra- and inter-mode correlations with the coupling strength for the two different types of coupling processes and analysed in details for three sets of states of the input modes. In the first, we will assume that both input modes are equally squeezed which corresponds to $n_{a}=n_{b}$ and $m_{a}=m_{b}$. In the second, we will assume that the input modes are equally populated but unequally squeezed which corresponds to $n_{a}=n_{b}$, $m_{a}\neq m_{b}$. In the third, we will assume that one of the input modes is in a squeezed state while the other is in the vacuum. 
Although the choice of state of the input modes affects the initial transient state of the output modes, we find that it also has effect on the steady-state of the output modes. Therefore, in most of the analysis we will focus our attention on the steady-state form of the variances, populations and the correlation functions.

\section{Linearly coupled modes}\label{sec3}

We first consider the effect of the linear coupling between the modes, described by the interaction Hamiltonian (\ref{e1}), on the variation of the fluctuations of the modes, their populations and one and two-photon correlations.
The conclusions will serve as reference for the case of the nonlinear coupling to follow.

Before proceeding, we would like to point out that there have been many studies of entanglement generation using the linear coupler, in particular a beam-splitter as the coupler of two modes~\cite{bs92,yb93,pa99,sw01,ks02,hf06,ti09,gt15,bb15}. It has been proved that at least one of the mode should be in a nonclassical state. In other words, at least one of the mode should contain nonclassical correlations. 

However, there are several questions not answered yet of how the two-mode nonclassical correlations are created from the existing single-mode nonclassical correlations? Are the two-mode correlations created by a transfer of the single-mode correlations or by conversion (transformation) of the single-mode correlations?
Here we will give answer to these questions identifying mechanisms responsible for the creation of the two-mode one-photon correlations, determining coherence properties of the modes, and the two-mode  two-photon correlations, necessary for entanglement.

Using the interaction Hamiltonian (\ref{e1}) in the Heisenberg-Langevin equations (\ref{e6}) we find that the quadrature components evolve as
\begin{align}
\dot{X}_{a} &= -\kappa X_{a} +g X_{b} - \sqrt{2\kappa}X^{\rm in}_{a}(t) ,\nonumber\\
\dot{X}_{b} &= -\kappa X_{b} -gX_{a} -\sqrt{2\kappa}X^{\rm in}_{b}(t) ,\nonumber\\
\dot{Y}_{a} &= -\kappa Y_{a} +g Y_{b} - \sqrt{2\kappa}Y^{\rm in}_{a}(t) ,\nonumber\\
\dot{Y}_{b} &= -\kappa Y_{b} -gY_{a} -\sqrt{2\kappa}Y^{\rm in}_{b}(t)  ,\label{e8}
\end{align}
from which we see that the dynamics of the $X$ quadratures are completely independent of the $Y$ quadratures. Equations~(\ref{e8}) can be solved by a direct integration, but we find that more convenient is to solve the equations introducing the Laplace transform of the quadrature operators and then solve the two pairs of algebraic equations, see Appendix for details.

\subsection{Variances of the quadrature components}\label{sec3a}

We begin by examining the effect of the linear coupling on the variances of the single-mode quadrature components of the modes which contain information on fluctuations properties of the modes. We focus on the evolution of the variances with time $t$ and the coupling strength $g$ when the input modes are unequally populated and their noise variances are unequally squeezed. Since the expectation values of the quadrature operators are all zero for Gaussian states, the variances of the quadrature components are just equal to the expectation values of the squares of the quadrature operators. In the Appendix explicit expressions for the quadrature components have been evaluated, Eq.~(\ref{ax}). Using these results and characterizing the coupling strength by a scaled variable (angle) $\psi =\arctan(g/\kappa)$, it is straightforward to derive expressions which show in a clear way the essential physics, we find that after an interaction time $t$ the variances of the modes are 
\begin{align}
\left\langle X^{2}_{a}(t)\right\rangle &= V_{+} +U_{-}\left[1-w_{l}(t,\psi) \right] ,\nonumber\\
\left\langle Y^{2}_{a}(t)\right\rangle & = V_{-} +U_{+}\left[1-w_{l}(t,\psi) \right] ,\nonumber\\ 
\left\langle X^{2}_{b}(t)\right\rangle &= V_{+} -U_{-}\left[1-w_{l}(t,\psi) \right] ,\nonumber\\
\left\langle Y^{2}_{b}(t)\right\rangle &=  V_{-} -U_{+}\left[1-w_{l}(t,\psi) \right]  ,\label{e9}
\end{align}
in which we have decomposed the variances into two parts 
\begin{align}
V_{\pm} = \left(\frac{1}{2}+n\pm m\cos^{2}\phi\mp \Delta m\sin^{2}\phi\right) ,\label{e11a}
\end{align}
containing contributions to the noise not affected by the interaction, and 
\begin{align}
U_{\pm} = \left(\Delta n\pm m\sin^{2}\phi\mp\Delta m\cos^{2}\phi\right) ,\label{e11b}
\end{align}
containing contributions affected by the interaction. For clarity of the expressions we have introduces the following notation
\begin{align}
n &= \frac{1}{2}(n_{a}+n_{b}) ,\ \Delta n=\frac{1}{2}(n_{a}-n_{b}) ,\nonumber\\
m &= \frac{1}{2}(m_{a}+m_{b}) ,\ \Delta m = \frac{1}{2}(m_{a}-m_{b}) .
\end{align}
The function
\begin{align}
w_{l}(t,\psi) = \sin\psi\left[\sin\psi - e^{-2\kappa t}\sin(2gt+\psi)\right] ,\label{e12w}
\end{align}
describes the modifications of the noise terms by the interaction. It contains the dependence of the variances on time $t$ and the coupling strength $\psi$.

As far as the effect of the strength of the coupling between the modes is concerned, the expression of the solution in terms of the angle $\psi$ leads to simple formulas for the variances which are easy for analysis and discussion. The results for the variances given by Eq.~(\ref{e9}) hold for an arbitrary state of the input modes which can vary from vacuum through unequally populated thermal states to unequally squeezed states. The $V_{\pm}$ terms contain parameters characterizing similarities of the input modes,  whereas terms $U_{\pm}$ contain parameters characterizing differences between the input modes, i.e. unequal populations of the modes $(\Delta n)$, unequal degrees of squeezing $(\Delta m)$ and different relative orientations of the noise ellipses $(\phi\neq 0)$. Clearly, the linear interaction between the modes affects only the differences between the modes, i.e., unequal population, unequal squeezing and unequal orientations of their noise ellipses. In other words, the coupling through the linear interaction can modify the variances only if the input modes are at least partly distinguishable in either the number of photons or degree of correlations or mutual orientation of the noise ellipses. If none of these conditions is fulfilled the interaction does not modify the variances.

 When the linear interaction effectively couples the modes the variances evolve in time and vary with the coupling strength. Specifically, the variation of the variances with $t$ and $\psi$ is determined by the function $w(t,\psi)$ given by Eq.~(\ref{e12w}). By inspection we observe that in the long time limit $w_{l}(t,\psi)$ reduces to $\sin^{2}\psi$. Hence, the factor $1-w_{l}(t,\psi)$ becomes independent of time and reduces to $\cos^{2}\psi$. From this simple analysis it follows that the second terms on the right-hand side of the variances (\ref{e9}) decrease with an increasing coupling strength and vanish for a very strong coupling at which $\psi =\pi/2$. In this limit $\langle X^{2}_{a}(t)\rangle = \langle X^{2}_{b}(t)\rangle = V_{+}$ and $\langle Y^{2}_{a}(t)\rangle = \langle Y^{2}_{b}(t)\rangle =V_{-}$, indicating that the linear interaction forces the fluctuations of the modes to become the same in both modes. In other words, the coupling forces the initially distinguishable modes through their variances to become indistinguishable. Since, in general,  $V_{+}\neq V_{-}$ the output modes are found in a squeezed state. 
 
 Since the values of the factors $V_{\pm}$ depend on the state of the input modes we see that the stationary state of the strongly coupled modes is critically dependent on the state of the input modes. 
 For example, the equally squeezed fields of the input modes $(\Delta n=0, \Delta m=0)$ with the relative phase $\phi=\pi/2$ are turned by the interaction to thermal states, $V_{\pm}=\frac{1}{2}+n$. The equally populated input modes but unequally squeezed are turned to equally weakly squeezed modes, $V_{\pm}=\frac{1}{2}+n\mp \Delta m$. Note that a quantum squeezed state, i.e. reduction of the variances below their vacuum level $\frac{1}{2}$ is possible in this case only when $\Delta m>n$.  
 
 If the field of one of the input modes, e.g. mode $a$ is squeezed and the field of the other is in vacuum, then the fluctuations of both modes are turned by the interaction to become equally squeezed,  $V_{\pm}=\frac{1}{2}\left(1+n_{a}\mp m_{a}\right)$. In this case reduction of $V_{+}$ below $\frac{1}{2}$, the limit for quantum squeezing is possible for $m_{a}>n_{a}$, i.e., when the field of the input squeezed mode is quantum squeezed.
 For maximally squeezed input mode, $m_{a}=\sqrt{n_{a}(n_{a}+1)}$, with $n_{a}\gg 1$ the squeezed variances of the output modes can be reduced up to $V_{+}=\frac{1}{4}$ indicating that under the interaction both output modes will be equally squeezed with the noise reduction up to $50\%$ below the vacuum limit.

 \subsection{Populations of the modes}
 
Similar condition to that observed for the variances applies for the populations of the modes that the interaction can modify the populations only if the modes were initially unequally populated. The variation of the populations with time and the linear coupling is found to be
\begin{align}
\langle a^{\dag}a\rangle  &= n+\Delta n\left[1 - w_{l}(t,\psi)\right] ,\nonumber\\
\langle b^{\dag}b\rangle &=  n-\Delta n\left[1- w_{l}(t,\psi)\right] .\label{e10}
\end{align}
We may discuss Eq.~(\ref{e10}) very briefly by remarking that the interaction modifies the populations only when the modes are unequally populated.
Note that the populations are independent of the two-photon correlations present inside the modes. The sum of the populations is constant and independent of the interaction as it should be since the linear interaction cannot produce photons. The interaction evidently tends to transfer the population from the more populated to less populated mode, i.e. the interaction also tends to equalise the populations. In other words, under a strong coupling the initially unequally populated modes $(\Delta n\neq 0)$ become equally populated. It is worth mentioning that the variation of the populations with $\psi$, the coupling strength, is formally identical with the variation of the variances of the quadratures of the mode operators.

\subsection{Single-mode correlations}

We now proceed to examine properties of the single-mode two-photon correlations. The time dependent solutions for the correlations are
\begin{align}
\langle aa\rangle &=  m_{a}e^{2i\phi} -\alpha m\sin\phi\, w_{l}(t,\psi)e^{i(\theta+\phi)} ,\nonumber\\
\langle bb\rangle &=  m_{b} +\alpha m\sin\phi\, w_{l}(t,\psi)e^{i(\theta+\phi)} ,\label{e11}
\end{align}
where
\begin{align}
\alpha = \sqrt{1+\cot^{2}\theta} ,\label{e12}
\end{align}
with
\begin{align}
\delta m =\frac{\Delta m}{m} ,\quad \cot\theta = \delta m\cot\phi , \quad \theta \in \left\langle0,\frac{\pi}{2}\right\rangle .\label{e12n}
\end{align}
It is already apparent by inspection of Eq.~(\ref{e11}) that the correlations may display quite different behaviour depending on whether the input modes were identically, $\delta m=0$, or non-identically, $\delta m\neq 0$, squeezed. In other words, action of the linear coupling on modes possessing equal two-photon correlations can be completely different than on modes possessing unequal correlations.

Let us examine these two cases in more detail. In practice one would prepare both input modes in equally squeezed states or one of the modes in a squeezed state and the other in a thermal or even in the vacuum state. 
If the input modes are initially in equally squeezed states $(\delta m=0)$, we see that in this case $\theta=\pi/2$ and then for optimal coupling $\phi=\pi/2$, we get that
\begin{align}
\langle aa\rangle &= -m\left[1- w_{l}(t,\psi)\right] .\nonumber\\
\langle bb\rangle &=  m\left[1- w_{l}(t,\psi)\right] .\label{e14a}
\end{align}
Apparently, the mutual coupling forces the correlations to gradually decrease as the strength of the coupling increases and to be suppressed altogether in the limit of a strong coupling.

This result can be understood by regarding the effect of the coupling strength on the variances of the modes. As we have seen the coupling tends to equalise variances of the input modes whose the squeezed noise ellipses are not parallel to each other. According to Eq.~(\ref{e8}) the interaction then couples the squeezed quadrature component of one of the modes with the unsqueezed component of the other mode. A strong coupling tends the variances of all four quadratures to become equal which is characteristic of a thermal field.

\begin{figure}[h]
  \includegraphics[width=0.49\textwidth]{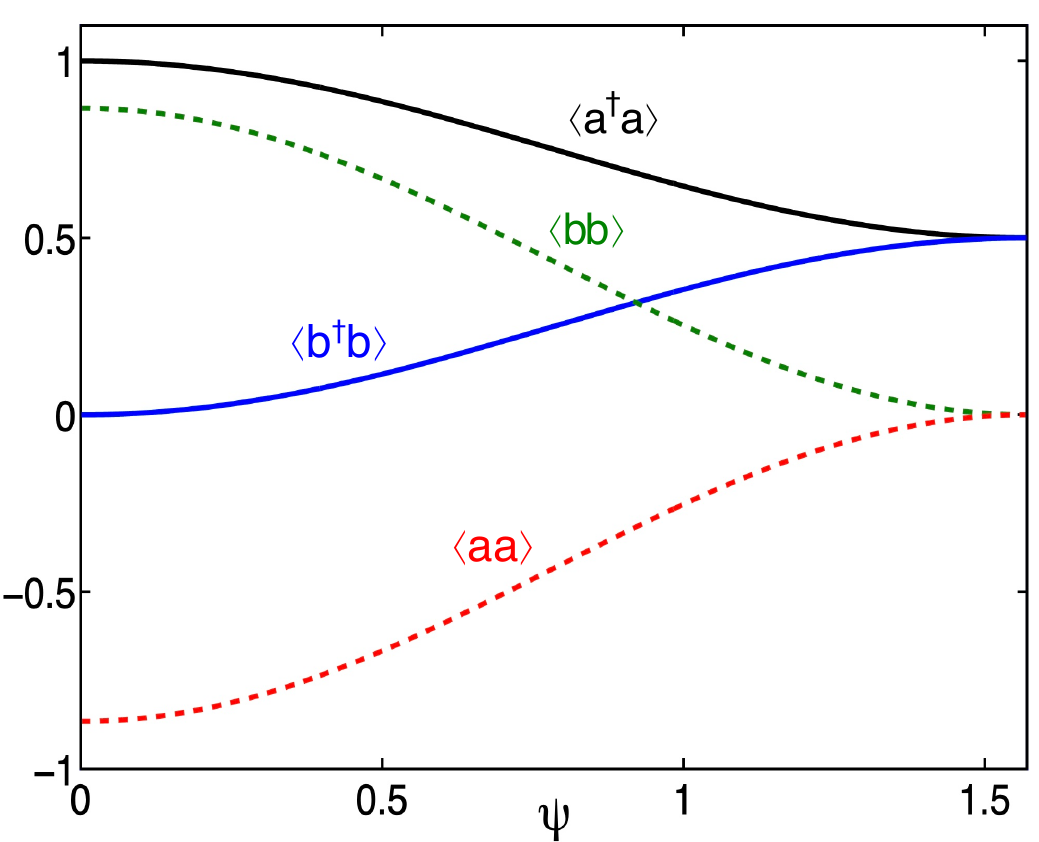}
  \caption{Variation of the populations and the single-mode two-photon correlations with $\psi$ for the case of the input mode $a$ being in a squeezed state with $n=0.5$ and $m=\sqrt{n(n+1)}$, and mode $b$ being in the vacuum.}
  \label{fig2a}
\end{figure}

The situation is quite different when one of the input modes, say mode $b$, is in uncorrelated state $(m_{b}=0)$. In this case, $\delta m=1$,  $m=m_{a}/2$, $\theta=\phi$, and then the correlations are given by
\begin{align}
\langle aa\rangle &=  m_{a} -\frac{1}{2}m_{a}w_{l}(t,\psi) ,\nonumber\\
\langle bb\rangle &=  \frac{1}{2}m_{a} w_{l}(t,\psi) ,\label{e11u}
\end{align}
In this case the correlations are seen to be transferred from mode $a$ to mode $b$ rather than being destroyed by the interaction. The major difference is that in this case the correlations tend to a nonzero value rather than to zero. It is clearly seen that up to one-half of the correlations present in the input mode $a$ can be transferred to mode $b$. Evidently, similar to the behaviour of the fluctuations and populations, the linear interaction tends to equalise two-photon correlations contained inside the modes. This is illustrated in Fig.~\ref{fig2a} where we plot the evolution of the two-photon correlations and the populations of the modes with the coupling strength $\psi$. It is seen that the effect of increasing the coupling strength is to equalise the correlations and the populations of the modes. For a very strong coupling $(\psi\approx \frac{\pi}{2})$ half of the population and correlations present in the input mode $a$ is transferred to mode $b$.

Further insight into variation of the two-photon correlations, in particular variation of their quantum behaviour is gained by considering the degrees of the correlations. These quantities are defined in Eq.~(\ref{e3}), and substituting Eqs.~(\ref{e10}) and (\ref{e11}) into Eq.~(\ref{e3}) yields the expressions
\begin{align}
\eta_{aa} &=  \frac{m}{n}\frac{\left|(1+\delta m)e^{2i\phi} -\alpha\sin\phi\,w_{l}(t,\psi) e^{i(\theta+\phi)}\right|}{1+\delta n\left[1 - w_{l}(t,\psi)\right]} ,\nonumber\\
\eta_{bb} &=  \frac{m}{n}\frac{\left|(1-\delta m) +\alpha\sin\phi\,w_{l}(t,\psi) e^{i(\theta+\phi)}\right|}{1-\delta n\left[1 - w_{l}(t,\psi)\right]} .\label{e14}
\end{align}
When we specialize Eq.~(\ref{e14}) to the case of equally populated $(\delta n =0)$ and equally squeezed fields $(\delta m=0)$ of the input modes, the degrees of the correlations inside the modes are identical and for $\phi =\frac{1}{2}\pi$ are
\begin{align}
\eta_{aa} &=  \eta_{bb} = \frac{m}{n}\left[1-w_{l}(t,\psi) \right] .\label{e14p}
\end{align}
From this it is clear that in the long time limit of $t\rightarrow \infty$ the degrees of the correlations decrease with the coupling strength $\psi$ from their input values $m/n$ to zero which is attained at $\psi=\pi/2$. Thus, independent of whether the modes are classically $(m<n)$ or quantum $(n<m\leq\sqrt{n(n+1)})$ squeezed, the interaction tends to destroy the correlations. 

\begin{figure}[h]
  \includegraphics[width=0.49\textwidth]{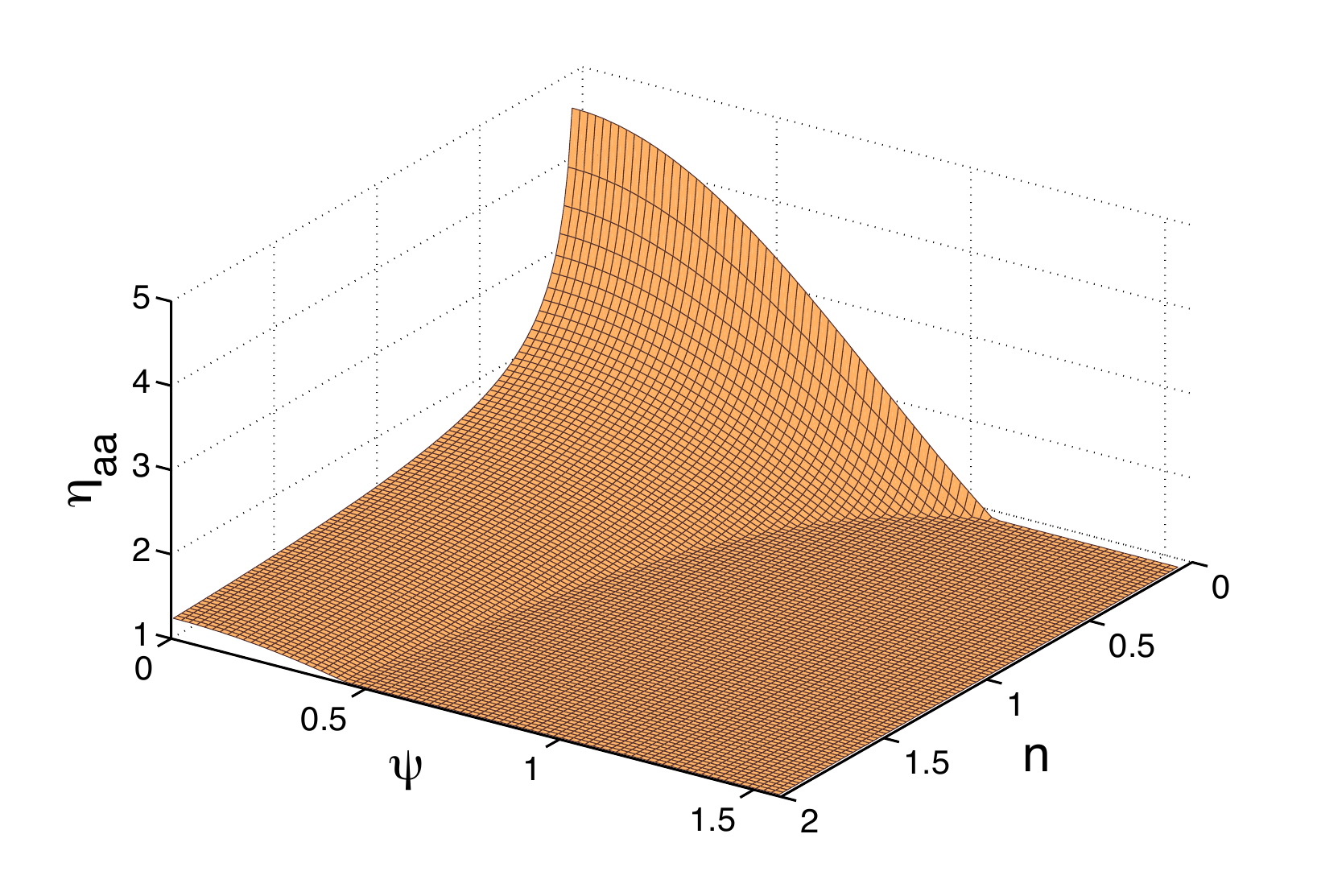}
  \caption{Variation of the single-mode two-photon correlations with $\psi$ and $n$ for both modes being in ideally squeezed state, $m=\sqrt{n(n+1)}$.}
  \label{fig3u}
\end{figure}

The rate at which the correlations are destroyed by the interaction depends not only on the coupling strength but also on the number of photons present in the modes. We may demonstrate this dependence by referring to the threshold value of the degree of the correlations $\eta_{ii}=1$ at which the correlations turn from quantum to classical. If we examine the expression (\ref{e14p}) we observe that condition  for the correlations to remain quantum $(\eta_{ii}>1)$  is that 
\begin{align}
\cos^{2}\psi > \sqrt{1-\frac{1}{n+1}} .\label{e15a}
\end{align}
which clearly illustrates that the range of $\psi$ over which the correlations remain quantum depends strongly on $n$. When the input modes contain a very small number of photons $(n\ll 1)$ the output modes will be in quantum squeezed states for almost the entire range of $\psi\in \langle 0,\frac{\pi}{2}\rangle$. On the other hand, for a large $n$ the right-hand side of Eq.~(\ref{e15a}) is close to one indicating that quantum squeezing of the modes is restricted to a very small range of $\psi$.  This is illustrated in Fig.~\ref{fig3u} where we plot the steady-state value of $\eta_{aa}$ as a function of $\psi$ and $n$ for ideally squeezed modes, i.e.,  $m=\sqrt{n(n+1)}$.

Quite different behaviour of the degrees of the correlations is observed when one of the input modes is in a squeezed state and the other in a thermal state of the same number of photons. Assume that the input mode $a$ was in a squeezed state with the degree of squeezing $m_{a}$ and number of photons $n_{a}$ but the input mode $b$ was in a thermal state of number of photons $n_{b}=n_{a}$. Then, $\delta m=1$, $\delta n=0$, $\theta=\phi$, and focussing of the long time limit, we find that the general expressions for the degrees, Eq.~(\ref{e14}), simplify to
\begin{align}
\eta_{aa} &=  \frac{m_{a}}{2n_{a}}\left(1+\cos^{2}\psi\right) ,\nonumber\\
\eta_{bb} &=  \frac{m_{a}}{2n_{a}}\sin^{2}\psi .\label{e22a}
\end{align}
Comparing this result with Eq.~(\ref{e14p}), we see that the behave of the degrees of the correlations is completely different than that for the case of both of the input modes being in a squeezed state. Currently, the interaction does not destroy the correlations. Since the sum $\eta_{aa}+\eta_{bb}=m_{a}/n_{a}$, we see that the degree of the correlations present in the input mode $a$ is preserved. Evidently, the interaction transfers the correlations between the modes without destroying them or converting into a different kind of correlations. In the limit of a very strong coupling $(\psi =\pi/2)$ the degrees of the correlations are identical.

\subsection{Two-mode correlations}

\begin{figure}[h]
  \includegraphics[width=0.49\textwidth]{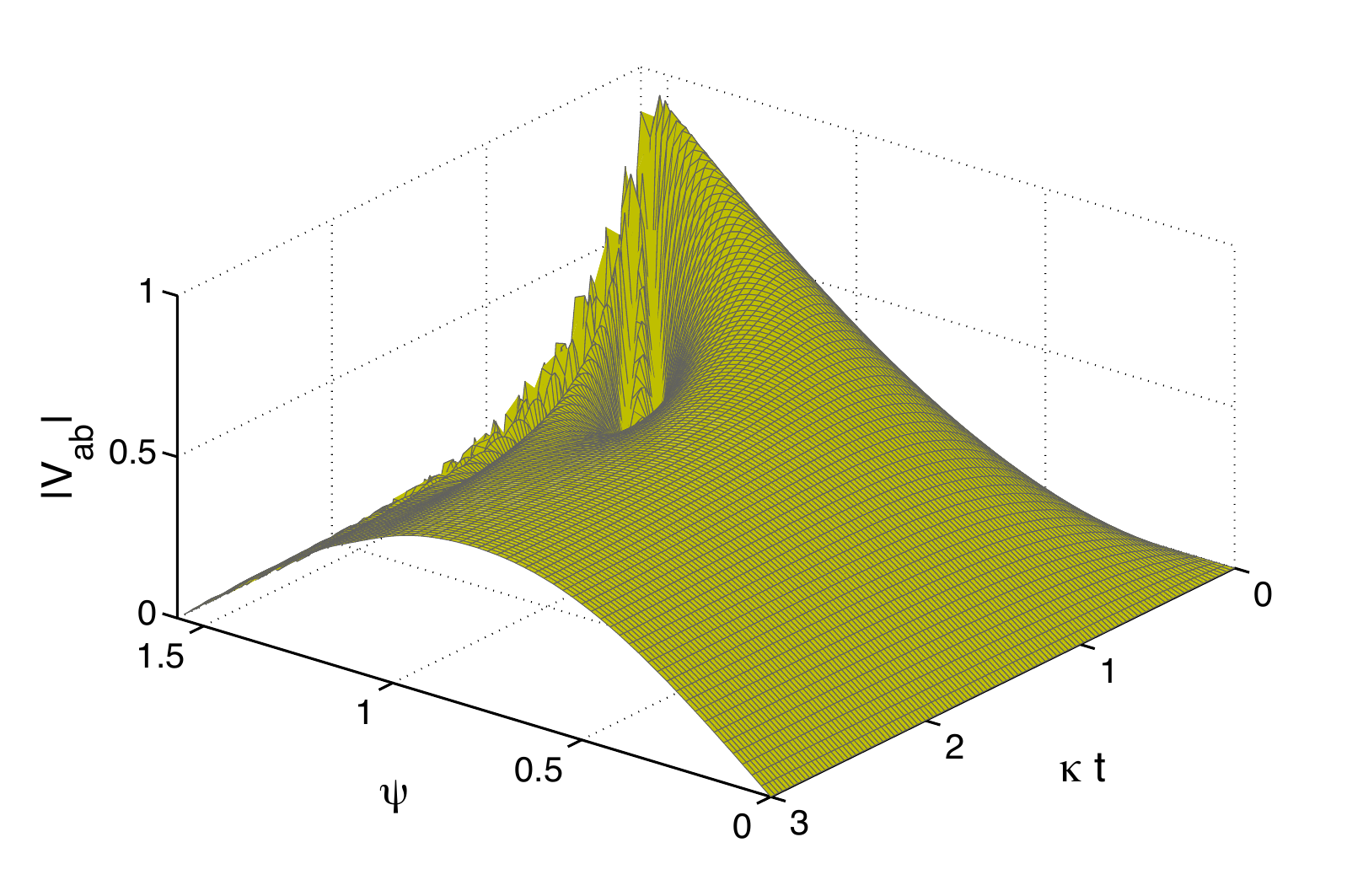}
  \caption{Variation of the visibility $|{\cal V}_{ab}|$ with the normalized time $\kappa t$ and $\psi$ for the case of the input mode $b$ being in the vacuum.}
  \label{fig3}
\end{figure}

So far our discussion of the correlation functions has been limited to single-mode correlations. However, the mutual interaction of the modes can create correlations between the modes. For Gaussian states two kinds of correlations can be created described by the cross correlation functions $\langle a^{\dag}b\rangle$ and $\langle ab\rangle$. In order to calculate $\langle a^{\dag}b\rangle$ and $\langle ab\rangle$ we make use of the solution of Eq.~(\ref{e8}).  Assuming that in general the input modes are in states of different numbers of photons $(n_{a}\neq n_{b})$ and different degrees of correlations $(m_{a}\neq m_{b})$ we find that the created by the interaction two-mode correlation functions are of the form
\begin{align}
\langle a^{\dag}b\rangle  &= -\Delta n\, u_{l}(t,\psi) ,\label{e22}
\end{align}
and
\begin{align}
\langle ab\rangle &=  -\alpha m\sin\phi\,u_{l}(t,\psi) e^{i(\theta+\phi)} ,\label{e23}
\end{align}
where
\begin{align}
u_{l}(t,\psi) = \sin\psi\left[\cos\psi - e^{-2\kappa t}\cos(2gt+\psi)\right] .\label{e25}
\end{align}
From Eq.~(\ref{e22}) it follows that the one-photon two-mode correlation function can be different from zero only if the input modes are unequally populated. Nonzero values of $\langle a^{\dag}b\rangle$ mean that the interaction turns unequally populated modes to be at least partly coherent. Thus, we expect to observe the first-order interference effects.
This result has a clear physical interpretation. It is reflection of the transport of the population from the more populated to less populated mode. For equally populated modes $\Delta n=0$ the correlations are zero since there is no transfer of photons between the modes. Regarding to the interference effect in this case it is possible, in principle, to determine by an examination of the modes from which of the two modes any photon was emitted. When $\Delta n\neq 0$ the interaction transfers photons from the more populated to less populated mode, which opens a second path for photons to reach a detector. There are, therefore, two probability amplitudes which interfere. When the interference occurs, we find that the interference pattern visibility is given by
\begin{align}
|{\cal V}_{ab}| &= \frac{2|\langle a^{\dag}b\rangle|}{\langle a^{\dag}a\rangle +\langle b^{\dag}b\rangle} = \left|\delta n\right| u_{l}(t,\psi) .\label{27u}
\end{align}
It is seen that the variation of the visibility follows the variation of $u(t,\psi)$, which results from  the action of the interaction. The maximum visibility is obtained for $|\delta n|=1$, which corresponds to the case of one of the input modes being in a thermal state and the other being in vacuum. 
The visibility given by Eq.~(\ref{27u}) is plotted in Fig.~\ref{fig3} as a function of normalized time $\kappa t$ and the coupling strength $\psi$. 
The visibility increases with $\psi$ and at short times can be close to unity when $\psi\approx \pi/2$. In the long time limit the maximum value of the visibility is $\frac{1}{2}$, which is attained at $\psi=\pi/4 \, (g=\kappa/2)$. Large values of the visibility at $\kappa t\ll 1$ are linked to the time-energy uncertainty relations. At short times the uncertainty of the energy (number of photons) of the modes is so large that it is impossible to predict from which mode photons are emitted.

Turning now to examine properties of the two-mode two-photon correlation function $\langle ab\rangle$, Eq.~(\ref{e23}), we first observe that $\langle ab\rangle$ can be nonzero only if at least one of the modes is in a squeezed state. Since
\begin{align}
\alpha\sin\phi = \sqrt{\sin^{2}\phi +(\delta m)^{2}\cos^{2}\phi} ,\label{e27}
\end{align}
we see that the variation of the correlation function with the relative phase $\phi$ depends crucially on whether the input modes are equally or unequally squeezed. If the modes are equally squeezed $(\delta m=0)$ the correlation varies with the phase as $\sin\phi$ function, indicating that it could vanish for $\phi =0$. If the modes are unequally squeezed $(\delta m\neq 0)$ the correlation function can be different from zero for any value of the phase $\phi$ and becomes independent of $\phi$ when $\delta m=1$, i.e. when one of the input modes in an uncorrelated (thermal or vacuum) state. 

Of greatest interest is the normalized two-photon correlation function $\eta_{ab}$, the degree of the two-photon correlations, since $\eta_{ab}>1$ is the necessary condition for entanglement. With help of Eqs.~(\ref{e10}) and (\ref{e23}) in Eq.~(\ref{e4}), we then readily obtain
\begin{align}
\eta_{ab} &=  \frac{m}{n}\,\frac{\alpha\sin\phi\, u_{l}(t,\psi)}{\sqrt{1-(\delta n)^{2}\left[1-w_{l}(t)\right]^{2}}} .\label{e28}
\end{align}
It is interesting to examine the degree of correlations predicted by this equation for various states of the input modes.
In particular, when $\delta n=0$, i.e., when the input modes were equally populated, we get
\begin{align}
\eta_{ab} &=  \frac{m}{n}\,\alpha\sin\phi\, u_{l}(t,\psi) ,\label{e28c}
\end{align}
which shows that there is no much difference in the variation of $\eta_{ab}$ with the coupling strength, determined by the function $u_{l}(t,\psi)$ when both input modes are equally populated. The degree of correlations varies with the coupling strength in the same manner independent of whether $\delta m =0\, (\theta =\pi/2)$ or $\delta m =1\, (\theta =\phi)$. 

It is easily verified that in the stationary limit the necessary condition for entanglement, $\eta_{ab} >1$, leads to an inequality
\begin{align}
\frac{m}{n}\alpha\sin\phi \sin\psi\cos\psi >1 .\label{e30}
\end{align}
Since $\alpha\sin\phi\sin\psi\cos\psi$ is always smaller than one, we see that this inequality cannot be satisfied when the input fields are in classically squeezed states, $m\leq n$,  but it can be satisfied when at least one of the input modes is in a quantum squeezed state, $m>n$. This result is in agreement with the conclusions made earlier in Refs.~\cite{ks02,hf06,ti09,gt15} that the linear interaction can entangle two modes only if the input modes were in quantum squeezed states.

When $\delta n\neq 0$, the degree of correlations varies with the coupling strength not only due to the variation of the correlations but also due to variation of the populations. In particular, when the field of the input mode $a$ is squeezed while the field of the input mode $b$ is in vacuum, $(\delta n=1)$, in the stationary limit the degree of the two-photon correlations is
\begin{align}
\eta_{ab} &=  \frac{m_{a}}{n_{a}}\frac{\cos\psi}{\sqrt{(2-\sin^{2}\psi)}} ,\label{e29}
\end{align}
which shows that the two-photon correlations can be created between squeezed and vacuum modes. Similar as in the previously discussed cases, in the strong coupling limit the correlations disappear.

\section{Nonlinearly coupled modes}\label{sec4}

In Sec.~\ref{sec3} we have investigated action of the linear coupling process on the fluctuations, populations, single- and two-mode correlations of two independent bosonic modes. 
The results have shown that the linear process effectively acts on the modes only if the modes are not identical that there is an asymmetry between the modes in either fluctuations, populations or two-photon correlations. When couples the modes the linear process turns them towards becoming identical, which can be achieved for a very strong coupling. In addition, we have established that the inter-mode correlations vary through a maximum attained at a moderate coupling strength, $g=\kappa/2$, and similar as the single-mode correlations vanish for a strong coupling.

In this section, we will contrast the action of the linear coupling with action of the nonlinear coupling on the modes. The starting point for our treatment are the Heisenberg-Langevin equations (\ref{e6}) which with the interaction Hamiltonian (\ref{e2}) take the form
\begin{align}
\dot{X}_{a} &= -\kappa X_{a} +g X_{b} - \sqrt{2\kappa}X^{\rm in}_{a}(t) ,\nonumber\\
\dot{X}_{b} &= -\kappa X_{b} +gX_{a} -\sqrt{2\kappa}X^{\rm in}_{b}(t) ,\nonumber\\
\dot{Y}_{a} &= -\kappa Y_{a} -g Y_{b} - \sqrt{2\kappa}Y^{\rm in}_{a}(t) ,\nonumber\\
\dot{Y}_{b} &= -\kappa Y_{b} -gY_{a} -\sqrt{2\kappa}Y^{\rm in}_{b}(t)  .\label{e35}
\end{align}
Simple comparison of Eqs.~(\ref{e35}) with the set of equations for the linear coupling, Eq.~(\ref{e8}), we see that these is a trivial difference between these two sets appearing only in the sign at $gX_{a}$ and $gY_{b}$. However, as we will see this trivial difference will lead to completely different solutions for the evolution of the quadratures with time and the coupling strength.

As in the previous section, we study action of the nonlinear interaction on the variances, populations, single-mode and two-mode correlations.

\subsection{Variances of the quadrature components}\label{sec4a}

The general solutions for the quadrature components are given in Eq.~(\ref{bx}). It is straightforward to evaluate the average values of squares of the quadrature components, which can be written in the form 
\begin{align}
\left\langle X^{2}_{a}(t)\right\rangle &= V_{+}\left[1+w_{n}(t,\chi) \right] + U_{-} ,\nonumber\\
\left\langle Y^{2}_{a}(t)\right\rangle & = V_{-}\left[1+w_{n}(t,\chi)\right] +U_{+} ,\nonumber\\ 
\left\langle X^{2}_{b}(t)\right\rangle &= V_{+}\left[1+w_{n}(t,\chi) \right] -U_{-} ,\nonumber\\
\left\langle Y^{2}_{b}(t)\right\rangle &= V_{-}\left[1+w_{n}(t,\chi) \right] -U_{+} ,\label{e36}
\end{align}
where $V_{\pm}, U_{\pm}$ are given in Eq.~(\ref{e11a}), and the dependence on time and the coupling strength is contained in the function
\begin{align}
w_{n}(t,\chi) = \sinh\chi\left[\sinh\chi - e^{-2\kappa t}\sinh(2gt+\chi)\right] ,\label{e38}
\end{align}
which is a function of time and the coupling strength expressed by a scaled variable
\begin{align}
\chi =\arctanh\left(\frac{g}{\kappa}\right) .\label{e39}
\end{align}
A number of comments could be made about the general solution for the variances presented in Eq.~(\ref{e36}). In the first place, we note that there is the crucial difference in the influence of the linear and nonlinear couplings on fluctuations of the modes. In the case of the nonlinear coupling the differences of the $X-X$ and $Y-Y$ variances are constant not affected by the coupling 
\begin{align}
\left\langle X^{2}_{a}(t)\right\rangle &- \left\langle X^{2}_{b}(t)\right\rangle = 2U_{-} ,\nonumber\\
\left\langle Y^{2}_{a}(t)\right\rangle &- \left\langle Y^{2}_{b}(t)\right\rangle = 2U_{+} ,\label{e38}
\end{align}
whereas the sums 
\begin{align}
\left\langle X^{2}_{a}(t)\right\rangle &+ \left\langle X^{2}_{b}(t)\right\rangle = V_{+}\left[1+w_{n}(t,\chi) \right]  ,\nonumber\\
\left\langle Y^{2}_{a}(t)\right\rangle &+ \left\langle Y^{2}_{b}(t)\right\rangle = V_{-}\left[1+w_{n}(t,\chi) \right]  ,\label{e39}
\end{align}
are enhanced by the nonlinear coupling. This means that the nonlinear coupling equally enhance the fluctuations in both modes without transferring the fluctuations between the modes. This is in contrast to the case of the linear coupling considered in Sec.~\ref{sec3}, where the interaction affected the difference of the variances leaving the sum unaffected. In that case the coupling transfers fluctuations between the modes without enhancing or diminishing fluctuations existing in the input modes. 

Secondly, we note that Eq.~(\ref{e36}) are physically different from those obtained for the linearly coupled modes, Eq.~(\ref{e9}). The main difference is that the function $w_{n}(t,\chi)$ grows rapidly with an increasing coupling strength and diverges to infinity when $g$ approaches the threshold value $g=\kappa\, (\chi\rightarrow\infty)$. This is in contrast to the linear coupling where the function $w_{l}(t,\psi)$ maximally can be equal to one. A consequence of the variation of $w_{n}(t,\chi)$ to infinity is that the variances increase rapidly to very large values indicating that the nonlinear coupling turns the modes to become highly fluctuation modes, so-called super-thermal fluctuations~\cite{bk85}.
Finally, we note that the solutions~(\ref{e36}) can be unstable. Rewriting the solutions (\ref{e36}) in terms of the coupling strength $g$ we can easily seen that there is a threshold value of $g=\kappa$ below which the solutions are stable. Since $\tanh\chi$ varies between $(-1,1)$, we see that writing the solutions in terms of the $\chi$ instead of $g$ ensures that the range of $\chi\in (0,\infty)$ solutions are stable. 

\begin{figure}[h]
  \includegraphics[width=0.49\textwidth]{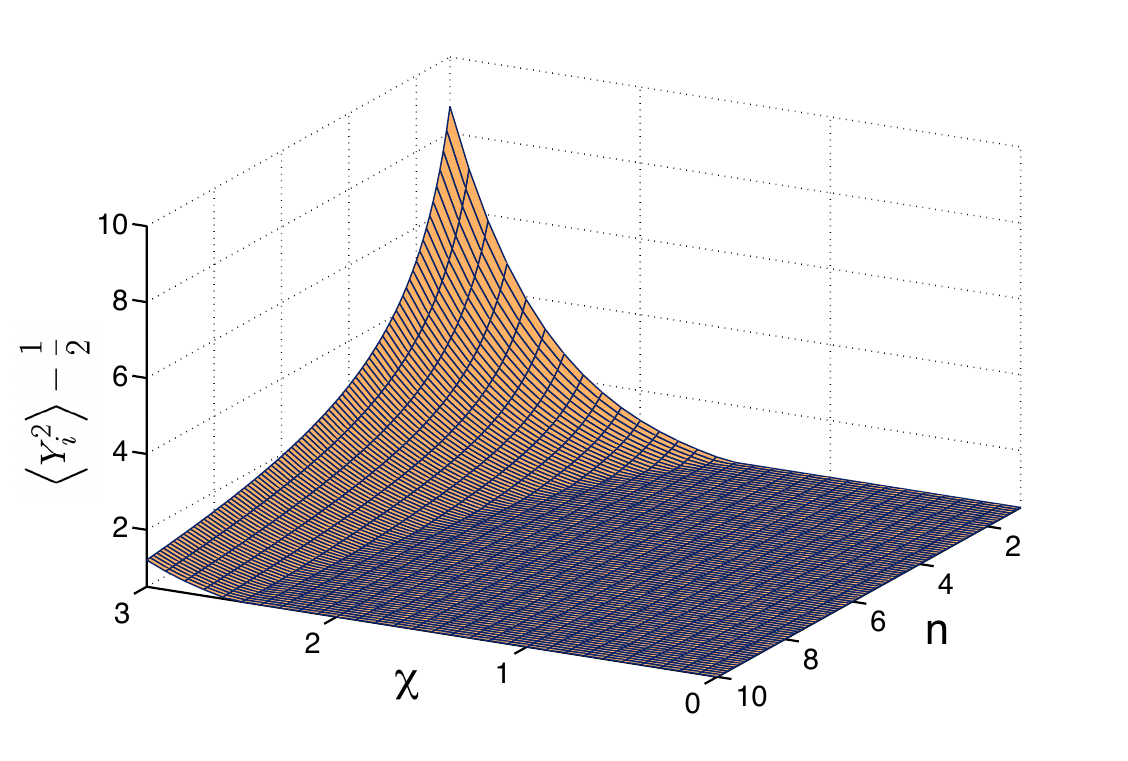}
  \caption{Variation of $\langle Y_{i}^{2}\rangle\, (i=a,b)$, the squeezed quadratures components, with $n$ and $\chi$ for $m=\sqrt{n(n+1)}$. }
  \label{fig5}
\end{figure}

The solution given by Eq.~(\ref{e36}) simplifies considerably in the special case of equally squeezed input modes $(\Delta n =0, \Delta m=0)$. In this case from Eq.~(\ref{e36}) we find that in long time limit 
\begin{align}
\left\langle X^{2}_{a}\right\rangle_{s} &= \left(\frac{1}{2}+n+m\cos^{2}\!\phi\right)\!\cosh^{2}\chi -m\sin^{2}\!\phi ,\nonumber\\
\left\langle Y^{2}_{a}\right\rangle_{s} &= \left(\frac{1}{2}+n-m\cos^{2}\!\phi\right)\!\cosh^{2}\chi + m\sin^{2}\!\phi ,\nonumber\\
\left\langle X^{2}_{b}\right\rangle_{s} &= \left(\frac{1}{2}+n+m\cos^{2}\!\phi\right)\!\cosh^{2}\chi +m\sin^{2}\!\phi ,\nonumber\\
\left\langle Y^{2}_{b}\right\rangle_{s} &= \left(\frac{1}{2}+n-m\cos^{2}\!\phi\right)\!\cosh^{2}\chi - m\sin^{2}\!\phi ,\label{e37}
\end{align}
where the subscript $s$ stands for the steady-state values.
It is clearly seen from Eq.~(\ref{e37}) that the nonlinear interaction equally amplifies those parts of the variances of the squeezed and anti-squeezed quadratures which are the same in both modes. The parts, $\pm\sin^{2}\phi$, which make a difference between variances of the $X_{a}$ and $X_{b}$ quadratures, and between the variances of the $Y_{a}$ and $Y_{b}$ quadratures are not affected by the coupling. Since $\cosh^{2}\chi$ groves nonlinearly with $\chi$, we have that the nonlinear interaction works as a nonlinear amplifier of the noise. For $\phi=0$ the interaction enhances the entire noise of the quadrature components whereas for $\phi=\pi/2$ the interaction enhances only a part of the noise arising from the vacuum and thermal fluctuations. Despite the fast groves of $\cosh^{2}\chi$ with $\chi$ the initially strongly quantum squeezed variances will remain quantum squeezed over a significant range of $\chi$. It is illustrated in Fig.~\ref{fig5}, where we have plotted the steady-state variances of the squeezed quadrature components $\langle Y_{i}^{2}\rangle$ as a function of $n$ and $\chi$. As the number of photons $n$ increases, the threshold value of $\chi$ at which the variances become larger than $\frac{1}{2}$ increases. despite the strength of the coupling $(\chi >2)$, the magnitude of the variances rapidly decrease with $n$.

\subsection{Populations of the modes}

The difference between the two types of coupling in affecting the variances of the modes is also reflected in the properties of the populations of the modes. The variations of the populations with time and with the coupling strength are
\begin{align}
\langle a^{\dag}a\rangle &= n_{a}  + \left(n+\frac{1}{2}\right)w_{n}(t,\chi) ,\nonumber\\
\langle b^{\dag}b\rangle &= n_{b} +\left(n+\frac{1}{2}\right)w_{n}(t,\chi) .\label{e41}
\end{align}
which shows that the nonlinear coupling equally builds up population of the modes independent of whether the input modes were equally or unequally populated. The builds up is from the fluctuations of the input modes not from their initial populations~\cite{mg67a,mg67b}. Thus, there is no transfer of the population between the modes but rather an equal build up of the population by the nonlinear coupling. It is easy to understand if one considers that the nonlinear interaction Hamiltonian indicates that every time a photon is created in one of the modes, simultaneously a photon is created in the other mode.

\subsection{Single-mode correlations}

Consider now how the nonlinear coupling influences on the two-photon correlations present in the input modes. From the solutions we find that the correlation functions at time $t$ and for the coupling strength $\chi$ are
\begin{align}
\langle aa\rangle  &= m_{a}e^{2i\phi} +\beta m\cos\phi\, w_{n}(t,\chi) e^{i(\phi+\theta)} ,\nonumber\\
\langle bb\rangle &= m_{b} +\beta m\cos\phi \, w_{n}(t,\chi)\, e^{-i(\phi+\theta)} ,\label{e42} 
\end{align}
and the corresponding degrees of the correlations are
\begin{align}
\eta_{aa} &= m\frac{\left|(1+\delta m)+\beta\cos\phi\, w_{n}(t,\chi)\, e^{i\left(\phi+\theta\right)}\right|}{ n(1+\delta n)+\left(n+\frac{1}{2}\right)w_{n}(t,\chi) } ,\nonumber\\
\eta_{bb} &= m\frac{\left|(1-\delta m)+\beta\cos\phi\, w_{n}(t,\chi)\, e^{-i\left(\phi+\theta\right)}\right|}{ n(1-\delta n)+\left(n+\frac{1}{2}\right)w_{n}(t,\chi) } .\label{e43}
\end{align}
where
\begin{align}
 \beta =\sqrt{1+\tan^{2}\theta} \ ,\qquad \tan\theta = \delta m\tan\phi .\label{e44}
\end{align}
The general expressions (\ref{e42}) show that the correlations are equally affected by the nonlinear coupling. It is worth noting that these correlation functions differ from that of the linearly coupled modes. As we have seen in Sec.~\ref{sec3}, the linear coupling led to a reduction of the correlations and to their disappearance in the limit of a very strong coupling. In the present case the situation is different. The nonlinear coupling enhances the existing correlations in the input modes or creates correlations in both modes even if one of the input modes was in uncorrelated state. To see it in more details consider some special cases. For example, when both input modes are equally squeezed $(\delta m=0, \theta=0, \beta =1)$, we get that the correlations depend on the phase $\phi$ and are maximal for $\phi=0$. This is in contrast to the case of linearly coupled modes where the correlations were maximal for $\phi=\frac{1}{2}\pi$. 

To examine the influence of the nonlinear coupling on quantum correlations present in the input modes we limit to the stationary case and readily find from Eq.~(\ref{e44}) that in the case of equally squeezed modes and $\phi=0$ the degrees are equal and vary with the coupling as
\begin{align}
\eta_{aa} &= \eta_{bb} = \frac{m}{n}\left[1- \frac{\sinh^{2}\chi}{2n+\left(2n+1\right)\sinh^{2}\chi}\right] .\label{e45}
\end{align}
Despite the fact that the nonlinear coupling enhances the two-photon correlations, the degrees of the two-photon correlations decrease with an increasing coupling strength. This is because the rate $(n+\frac{1}{2})$, at which the populations are increasing exceeds the rate $m$ at which the correlations increase. It is easy to see, the maximal value of the correlations $m=\sqrt{n(n+1)}$ is always smaller than $(n+\frac{1}{2})$ and approach $(n+\frac{1}{2})$ in the limit of $n\rightarrow\infty$. 

Interesting behaviour of the nonlinear coupling can be found by taking the limit of a strong coupling, $\chi\rightarrow\infty$. In this limit Eq.~(\ref{e45}) directly shows that the correlations do not decay to zero. The nonlinear coupling has tendency to reduce quantum correlations present in the input modes to the level of maximal classical correlations.
This is in contrast to the case of linearly coupled modes where in the limit of a strong coupling the correlations were reduced to zero, i.e. the correlated states of the input modes were turned to thermal states. 

Finally, we consider the situation when two-photon correlations are present only in one of the input modes. For example, if the correlations are present in the input mode $a$, we find that the nonlinear coupling induces two-photon correlations in mode $b$. The amount of the induced correlations depends on the state of mode $b$. If mode $b$ was in thermal state of the same number of photons present in mode $a$, then the degree of the correlations for mode $b$ is
\begin{align}
\eta_{bb} = \frac{mw_{n}(t,\chi)}{2n+\left(2n+1\right)w_{n}(t,\chi) } .\label{e47}
\end{align}
When mode $b$ was in the vacuum state then 
\begin{align}
\eta_{bb} =  \frac{mw_{n}(t,\chi)}{\left(n+1\right)w_{n}(t,\chi) } .\label{e48}
\end{align}
It is not difficult to verify that the degree of correlations induced in mode $b$ cannot excise the threshold value for quantum correlations, $\eta_{bb}=1$. The asymptotic form of Eq.~(\ref{e47}) for $\chi\rightarrow\infty$  and $m=\sqrt{n(n+1)}$ is
\begin{align}
\eta_{bb} = \frac{1}{2}\sqrt{1-\frac{1}{(2n+1)^{2}}} ,\label{e49}
\end{align}
from which we see that the maximum value is achieved when $n\rightarrow\infty$, in which case $\eta_{bb}=\frac{1}{2}$. The asymptotic form of Eq.~(\ref{e48}) for $\chi\rightarrow\infty$  and $m=\sqrt{n(n+1)}$ is
\begin{align}
\eta_{bb} = \sqrt{\frac{n}{(n+1)}} ,\label{e50a}
\end{align}
from which we see that the maximum value is reached when $n\rightarrow\infty$, in which case $\eta_{bb}=1$. It follows that in the limit $\chi\rightarrow\infty$ the two modes become in effect classically correlated. We may conclude that even if quantum correlations are present in one of the input modes, the nonlinear coupling between the modes can convert them only to the perfect classical correlations in the other mode.

\subsection{Inter-mode correlations}

We now turn to investigate properties of the inter-mode one- and two-photon correlations which can be generated by the nonlinear coupling process. These two types of correlations are determined by the correlation functions which at time $t$ and for the coupling strength $\chi$ are 
\begin{align}
\langle a^{\dag}b\rangle &= \beta m\cos\phi\, u_{n}(t,\chi)e^{-i(\theta+\phi)} ,\label{e50}
\end{align}
and
\begin{align}
\langle ab\rangle &= \left(n+\frac{1}{2}\right) u_{n}(t,\chi)  ,\label{e51}
\end{align}
in which
\begin{align}
u_{n}(t,\chi) = \sinh\chi\left[\cosh\chi -e^{-2\kappa t}\cosh(2gt+\chi)\right] .\label{e52}
\end{align}
The function $u_{n}(t,\chi)$ determines variation of the correlation functions with time $t$ and with the coupling strength $\chi$. It is clear seen that the function differs from $w_{n}(t,\chi)$ which determines variation of the populations and the single-mode correlations. 
\begin{figure}[h]
  \includegraphics[width=0.49\textwidth]{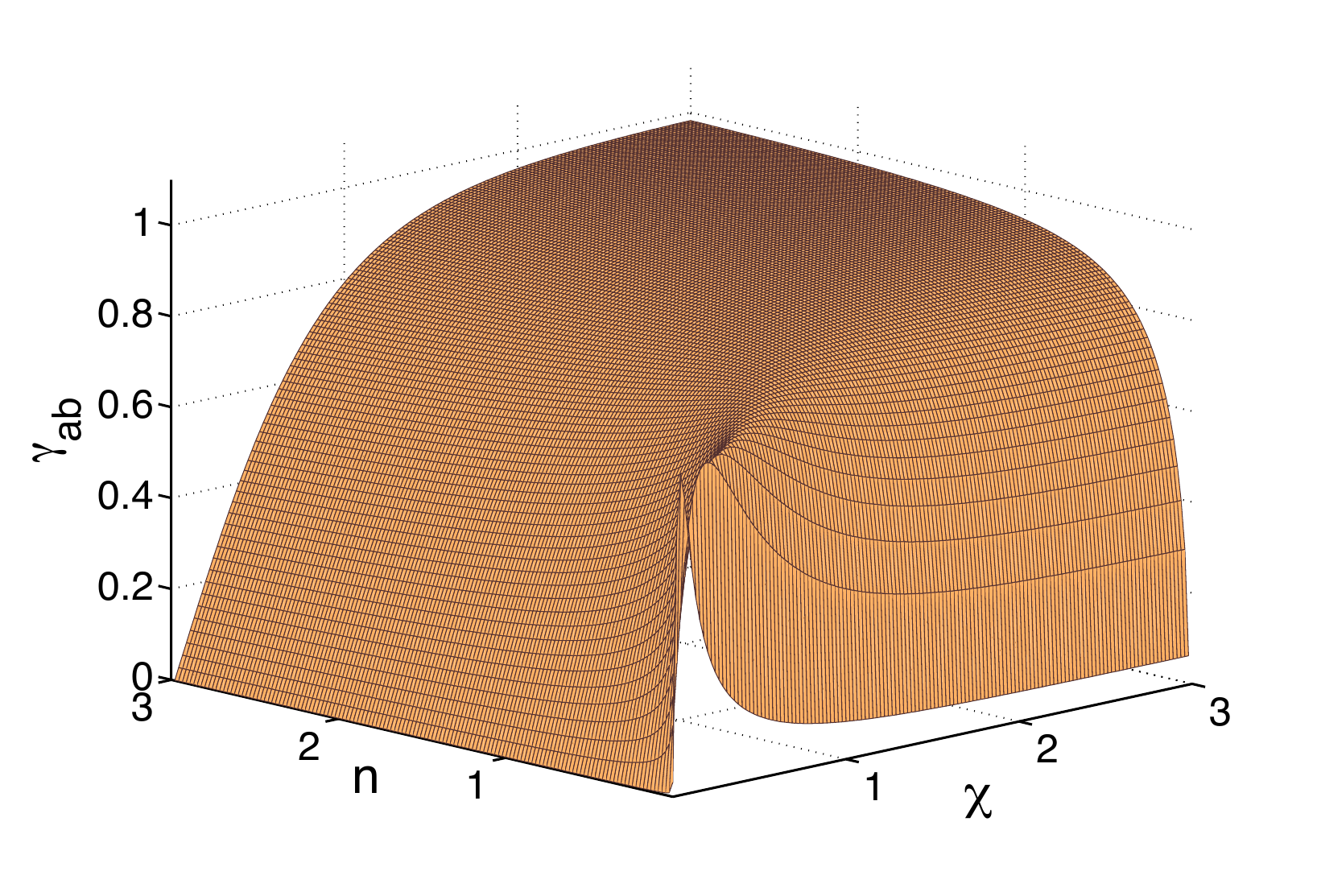}
  \caption{Variation of the degree of the one-photon coherence $\gamma_{ab}$, the first-order coherence,  with the coupling strength $\chi$ and the number of photons $n$ for equally squeezed input modes with the maximal quantum correlations $m=\sqrt{n(n+1)}$. }
  \label{fig6}
\end{figure}

Certain general features of the correlation functions follow from Eqs.~(\ref{e50}) and (\ref{e51}). The correlation functions are very different in forms compared to those found in Sec.~\ref{sec3} for the linearly coupled modes. In particular, the single-mode two-photon correlations $m$ present in the input modes are not themselves involved in the creation of the inter-mode two-photon correlations. The correlations are involved in the creation of the first-order correlations between the modes. It is quite surprising to find that the two-photon correlations are involved in the creation of the two-mode one-photon correlations instead of what one could expect that the correlations should be involved in the creation of the two-mode two-photon correlations.

Let us focus on the degrees of the inter-mode correlations, in particular the degree of the two-photon correlations $\eta_{ab}$, which provides information about entanglement between the modes. With the help of Eqs.~(\ref{e41}), (\ref{e50}) and (\ref{e51}), the degrees of the correlations are
\begin{align}
\gamma_{ab} &= \frac{2\beta\,  m\, u_{n}(t,\chi)|\cos\phi|}{\sqrt{\left[2n+\left(2n+1\right)w_{n}(t,\chi)\right]^{2} -4n^{2}(\delta n)^{2}}} ,\nonumber\\
\eta_{ab} &=  \frac{ \left(2n+1\right)u_{n}(t,\chi)}{\sqrt{\left[2n+\left(2n+1\right)w_{n}(t,\chi)\right]^{2} -4n^{2}(\delta n)^{2}}} .\label{e53}
\end{align}
If the input modes were equally squeezed, $\delta n=0, \phi=0$, the degrees of the correlations reduce to
\begin{align}
\gamma_{ab} &= \frac{2m u_{n}(t,\chi)}{2n+\left(2n+1\right)w_{n}(t,\chi)} ,\label{e54}\\
\eta_{ab} &=  \frac{(2n+1)u_{n}(t,\chi)}{2n+\left(2n+1\right)w_{n}(t,\chi)} .\label{e55}
\end{align}
It is easily verified that in this case the visibility is $|{\cal V}_{ab}|=\gamma_{ab}$.

Figure~\ref{fig6} shows the variation of the first-order coherence with the coupling strength $\chi$ and the number of photons $n$. The coherence increases with the coupling strength and with the number of photons and can be equal to unity indicating that in this limit the modes become perfectly coherent. The most interesting aspect of the first-order coherence shown in Fig.~\ref{fig6} relates to its insensitivity to the number of photons present in the input modes because this feature indicates that the modes will be perfectly coherent even at large temperatures. The dependence of the one-photon coherence on $n$ is a feature completely opposite to that exhibited by the two-photon coherence, which retains its nonclassical behaviour only at small $n$. At large $n$ the quantum correlations are reduced to maximal classical correlations.
\begin{figure}[h]
  \includegraphics[width=0.49\textwidth]{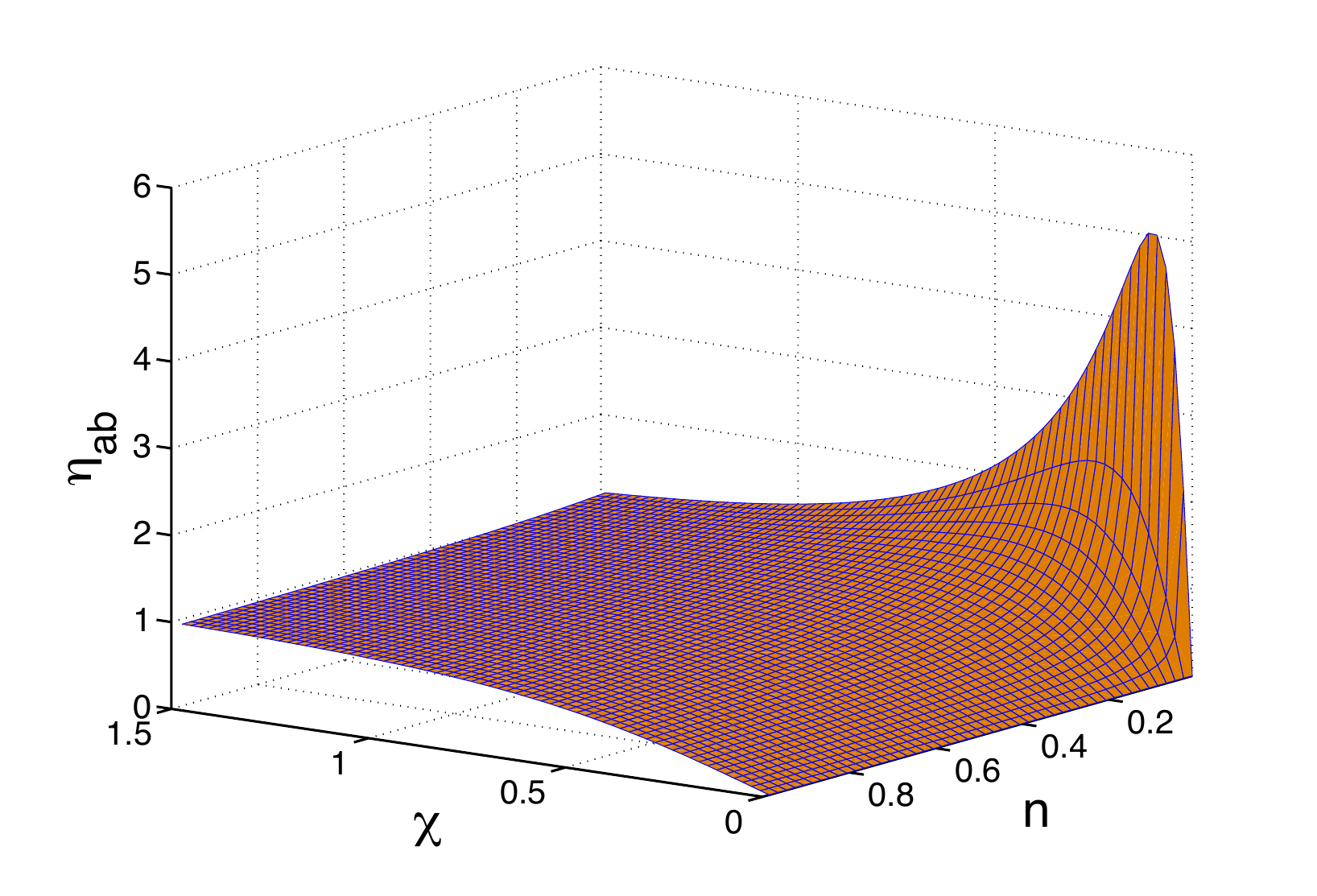}
  \caption{Variation of the degree of the inter-mode two-photon correlations $\eta_{ab}$ with the coupling strength $\chi$ and its dependence on the number of photons $n$ for equally squeezed input modes with the maximal single-mode quantum correlations $m=\sqrt{n(n+1)}$. }
  \label{fig7}
\end{figure}

Figure~\ref{fig7} shows the corresponding situation for the degree of the two-photon correlations. As a function of the coupling strength the correlations show different behaviour in that for small $n$ the correlations rapidly increase above the threshold value $\eta_{ab}=1$ and rapidly decline thereafter. For large $n$ the correlations smoothly increase to the maximal value one. Thus at large $n$ and for strong couplings the modes exhibit classical two-photon correlations. This is in contrast to the linearly coupled modes which for strong couplings are completely uncorrelated.

\section{Connection between single and two-mode correlations}\label{sec5}

When investigating the variation of the correlations with the coupling strength, we have noticed that the manner in which the correlations vary with the coupling strength is different for the single- and two-mode correlations. The variation of the single-mode correlations, as well as the populations of the modes, for the linear and nonlinear couplings is determined, respectively, by the functions $w_{l}(t,\psi)$ and $w_{n}(t,\chi)$, whereas the variation of the two-mode correlation functions is determined, respectively, by the functions $u_{l}(t,\psi)$ and $u_{n}(t,\chi)$. For the linear coupling the stationary value of the function $w_{l}(t,\psi)$ varies with the coupling strength $\psi$ as $\sin^{2}\psi$, while $u_{l}(t,\psi)$ varies as $\sin\psi\cos\psi$. For the nonlinear coupling the the stationary value of the function $w_{n}(t,\chi)$ varies with the coupling strength $\chi$ as $\sinh^{2}\psi$, while $u_{n}(t,\chi)$ varies as $\sinh\chi\cosh\chi$.

A question then arises concerning the connection between variation of the single mode populations and correlations, and the creation of the two-mode correlations. Therefore, in this section we will search for that connection.

In the case of the linearly coupled modes, a close look at the one-photon correlation function $\langle a^{\dag}b\rangle$, Eq.~(\ref{e22}), and the populations of the modes, Eq.~(\ref{e10}), suggests that it is possible, in principle, to associate the stationary value of the one-photon correlation function $\langle a^{\dag}b\rangle$ with the variation with the coupling strength $\psi$ of the stationary population of either mode $a$ or mode $b$ that 
\begin{align}
\left|\left\langle a^{\dag}b\right\rangle_{s}\right| = \left|\frac{1}{2}\frac{\partial}{\partial \psi}\langle a^{\dag}a\rangle_{s}\right|  ,\label{e56}
\end{align}
where the subscript $s$ stands for the stationary value. Evidently, the two-mode one-photon correlation function is directly related to the variation of the populations of the modes with the coupling strength $\psi$. Hence, we may interpret this relation that the one-photon correlations arise from the variation of the population of the modes with the coupling strength $\psi$. In other words, variation of the population of the modes with the coupling strength results in the creation of the two-mode one-photon correlation function.

Similarly, by comparing the two-photon correlation function $\langle ab\rangle$, Eq.~(\ref{e23}), with the single-mode two-photon correlation function of either mode $a$ or mode $b$, we find that
\begin{align}
\left|\left\langle ab\right\rangle_{s}\right| = \left|\frac{1}{2}\frac{\partial}{\partial \psi}\left\langle aa\right\rangle_{s}\right| .
\end{align}
Thus, the two-photon correlations can be associated with the variation of the single-mode two-photon correlations.
\begin{figure}[h]
 \includegraphics[width=0.43\textwidth]{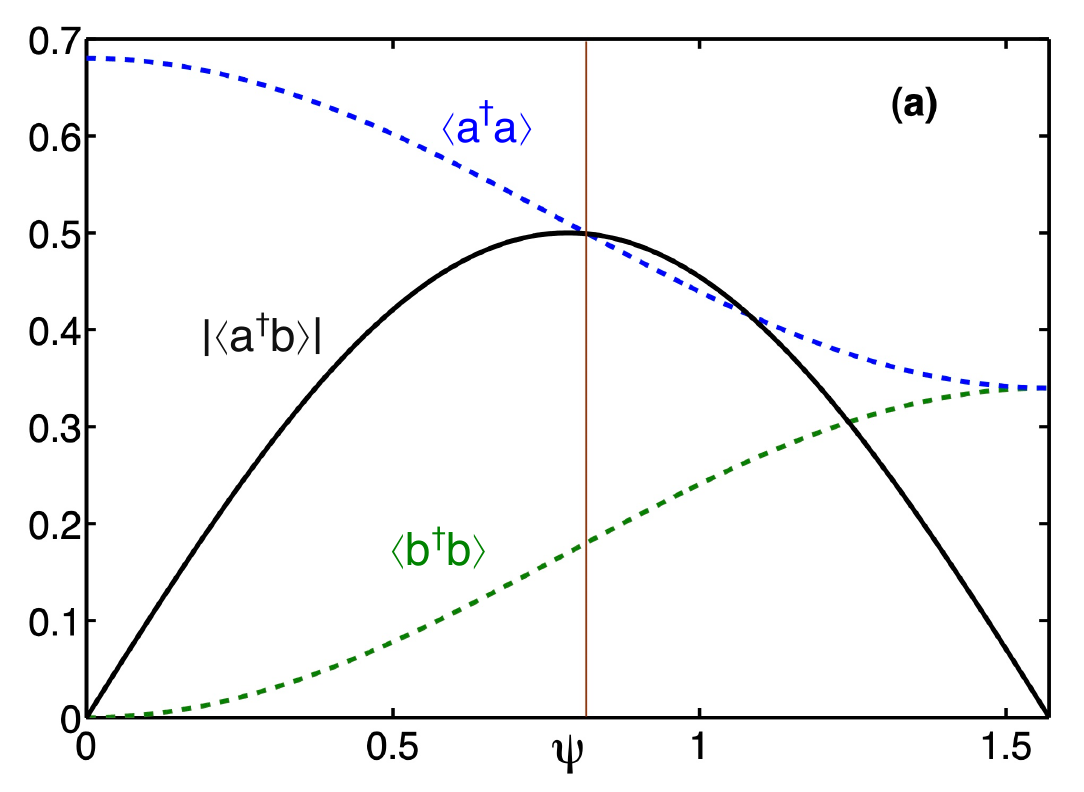}\\
  \includegraphics[width=0.5\textwidth]{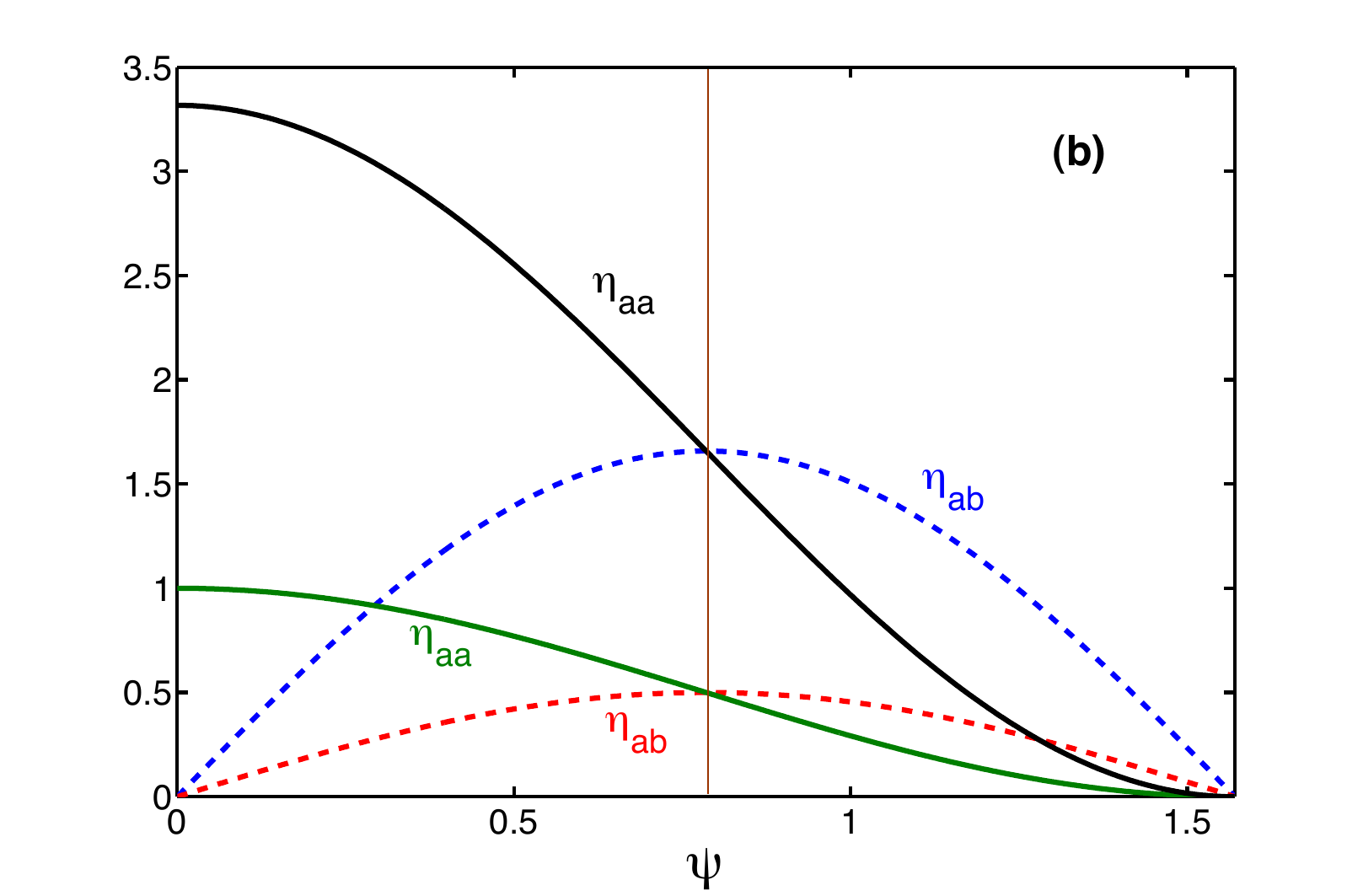}
  \caption{(a) Variation of the populations of the modes and the one-photon correlation function with $\psi$ for $\phi=\pi/2$ and $n=0.1$. Maximum of $|\langle a^{\dag}b\rangle|$ is at the inflection point of the populations whose the position is indicated by the vertical red line. (b) Variation of the degrees of the two-photon correlations with $\psi$ for $\phi=\pi/2$ and $n = 0.1$ and two different degrees of the single-mode two-photon correlations; $m=\sqrt{n(n+1)}$ (solid black line) and $m=n$ (solid green line).  Maximum of $\eta_{ab}$ is at the inflection point of $\eta_{aa}$.}
  \label{fig8}
\end{figure}

In order to support the above relations, we plot in Fig.~\ref{fig8} the variation of the two-mode correlation functions with the coupling strength $\psi$ for $\phi=\pi/2$ and $n=0.1$. 
Figure~\ref{fig8}(a) shows the variation of the populations of the modes and the correlation function $|\langle a^{\dag}b\rangle|$. It is apparent that the correlations arise from the variation of the populations. The magnitude of the correlations depends on how fast the populations vary with the coupling strength and it is clearly seen that the maximum of the correlation is at the inflection points of the varying populations, i.e., the maximum of the correlations is at the point in which the variation of the populations is the fastest.
In Fig.~\ref{fig8}(b) we compare variation of the two-mode two-photon correlations with the variation of the single-mode two-photon correlations. We have plotted the degrees of the correlations, i.e., the correlations divided by the number of photons to show that the relation between the correlations holds in regime of both classical and quantum correlations. Similar to the one-photon correlations, the two-photon correlations are maximal at the inflection point of the single-mode two-photon correlations. 

Thus, we may conclude that the linear interaction between the modes creates one-photon correlations (coherence) between the modes through the variation with the coupling strength of the populations of the modes. Similarly, the two-mode two-photon correlations are created through the variation with the coupling strength of the single-mode two-photon correlations.

Similar connections between the single- and two-mode correlations can be derived for the case of nonlinearly coupled modes. By comparing the two-mode one-photon correlations, Eq.~(\ref{e50}), with the single-mode two-photon correlations, Eq.~(\ref{e42}) we may write that
\begin{align}
\left|\left\langle a^{\dag}b\right\rangle_{s}\right| = \left|\frac{1}{2}\frac{\partial}{\partial \chi}\langle aa\rangle_{s}\right|  .\label{e58}
\end{align}
and a comparison of the two-mode two-photon correlations, Eq.~(\ref{e51}) with the populations of the modes, Eq.~(\ref{e41}) leads to the relation
\begin{align}
\left|\left\langle ab\right\rangle_{s}\right| = \left|\frac{1}{2}\frac{\partial}{\partial \chi}\left\langle a^{\dag}a\right\rangle_{s}\right| .
\end{align}
Clearly, the creation of the inter-mode  two-photon correlations is intimately connected to variation of the amplification of the population of the modes.
Hence, we may conclude that the creation of the inter-mode one-photon correlations results from the variation with the coupling strength $\chi$ of the single-mode two-photon correlations, whereas the inter-mode two-photon correlations arise from the variation of the populations of the modes with the coupling strength $\chi$. 

In summary of this section, we have seen that in order to generate the inter-mode correlations one must not only prepare the modes in two-photon states, but also the populations and the two-photon correlations should vary with the coupling strength.

\section{Summary}\label{sec6}

We have studied the problem of the creation of inter-mode correlations by application of two different types of coupling processes between two Gaussian bosonic modes; the linear and nonlinear two-photon coupling processes. Two kinds of the inter-mode correlations have been studied, the one-photon correlation function $\langle a^{\dag}b\rangle$, which carries information about coherence properties of the modes, and the two-photon correlation function $\langle ab\rangle$, which reveals a quantum feature in that it is necessary for entanglement between the modes. We have also investigated the ways the two types of coupling processes affect the internal states of the modes. We have presented general analytic formulas for the inter-mode correlations as well as for the mode populations and two-photon correlations valid for an arbitrary time $t$ and arbitrary coupling strengths, which in the case of the nonlinear process are restricted by the stability conditions imposed of the solutions. We have assumed that the modes are in squeezed states which are not necessary equally squeezed or having equally oriented noise ellipses.  

We have found significant differences in the conditions required by these two processes to effectively couple the modes. The linear process effectively couples the modes only when there are differences in populations of the modes or two-photon correlations or in the mutual orientation of the noise ellipses. On the other hand, the nonlinear process effectively couples the modes independent of any differences in the population and correlation properties of the modes. Our investigations show that the linear process tends to diminish initial differences between the modes to wash them in the limit of a strong coupling. Depending on the initial populations and two-photon correlations the modes can be turned by the interaction to evolve towards thermal states of the same numbers of photons or towards equally squeezed states. Regarding the inter-mode correlations we have found that the one-photon correlation function is created under the condition that the modes are unequally populated. The creation of the inter-mode two-photon correlations requires a difference in the amount or redistribution of the single-mode two-photon correlations.

The nonlinear process amplifies the populations of the modes and that part of the noise variances which arise from the contribution of the vacuum and thermal fluctuations.
Two-photon correlations present in the modes can be amplified only when the modes are equally squeezed.
Despite the fact the the single-mode two-photon correlations are amplified by the interaction the degree of the two-photon correlation decreases with the increasing coupling strength and tends towards unity, the value corresponding to the maximally correlated classical states. The inter-mode two-photon correlations are  generated independent of the two-photon correlations present in the modes. We have found that the interaction turns the two-photon correlations into the inter-mode one-photon correlations such that in  
the strong coupling limit the modes are found to be perfectly coherent with the degree of coherence equal unity. In addition, the modes are left in correlated two-photon states with the degree of correlation corresponding to the maximally correlated classical states.  

Finally, we have shown that there is a simple connection between the single mode populations and correlations and the creation of the two-mode correlations.
Our analysis has led to the conclusion that the generation of the inter-mode correlations results from a variation of the populations and single-mode two-photon correlations with the coupling strength. 

The results presented here can be used as the basis for the study of correlation properties of large numbers of modes arranged in one or two dimensional quantum networks.

\begin{acknowledgments}
	We would like acknowledge funding by the Minister of Science under the "Regional Excellence Initiative"  program, Project No. RID/SP/0050/2024/1.

\end{acknowledgments}

\appendix

\section{}

In this appendix we give the solutions of the Heisenberg-Langevin equations for the linear and nonlinear coupling processes, and expressions for the populations and the correlation functions.

\subsection{Solution of the Heisenberg-Langevin equations for the linear coupling}\label{seca1}

The Laplace transform of the set of coupled differential equations for the quadratures components $X_{i}$ and $Y_{i}$ under the linear coupling process, Eq.~(\ref{e8}), leads to two sets of equations

\begin{align}
\left(
\begin{array}{cc}
p+\kappa & g  \\
-g & p+\kappa
\end{array}
\right) \left(
\begin{array}{c}
X_{b}(p)\\
X_{a}(p)
\end{array}
\right) = \left(
\begin{array}{c}
L_{bx}(p)\\
L_{ax}(p)
\end{array}
\right) 
\end{align}
and
\begin{align}
\left(
\begin{array}{cc}
p+\kappa & -g  \\
g & p+\kappa
\end{array}
\right) \left(
\begin{array}{c}
Y_{a}(p)\\
Y_{b}(p)
\end{array}
\right) = \left(
\begin{array}{c}
L_{ay}(p)\\
L_{by}(p)
\end{array}
\right) 
\end{align}
where
\begin{align}
L_{ax}(p) &= \left[X_{a}(0) -\sqrt{2\kappa}X_{a}^{\rm in}(p)\right] ,\nonumber\\
L_{ay}(p) &= \left[Y_{a}(0) -\sqrt{2\kappa}Y_{a}^{\rm in}(p)\right] ,\nonumber\\
L_{bx}(p) &= \left[X_{b}(0) -\sqrt{2\kappa}X_{b}^{\rm in}(p)\right] ,\nonumber\\
L_{by}(p) &= \left[Y_{b}(0) -\sqrt{2\kappa}Y_{b}^{\rm in}(p)\right] .
\end{align}
Solving for $X_{b}(p)$ and $X_{a}(p)$ we find
\begin{align}
X_{b}(p) &= \frac{(p+\kappa)L_{bx}(p) -gL_{ax}(p)}{(p-p_{1})(p-p_{2})} , \nonumber\\
X_{a}(p) &= \frac{(p+\kappa)L_{ax}(p) +gL_{bx}(p)}{(p-p_{1})(p-p_{2})} , \label{a4}
\end{align}
and for  $Y_{a}(p)$ and $Y_{b}(p)$:
\begin{align}
Y_{a}(p) &= \frac{(p+\kappa)L_{ay}(p) +gL_{by}(p)}{(p-p_{1})(p-p_{2})} ,\nonumber\\
Y_{b}(p) &= \frac{(p+\kappa)L_{by}(p) -gL_{ay}(p)}{(p-p_{1})(p-p_{2})}  .\label{a5}
\end{align}
where 
\begin{align}
p_{1}= -\kappa +ig ,\qquad p_{2} = -\kappa -ig .
\end{align}
Taking the inverse Laplace transforms, we find that the time evolutions of the quadratures are
\begin{align}
X_{b}(t) &= \frac{1}{2}\left\{[L_{bx}(p_{1})+iL_{ax}(p_{1})]e^{p_{1}t}\right. \nonumber\\
&\left. \qquad  +[L_{bx}(p_{2})-iL_{ax}(p_{2})]e^{p_{2}t} \right\} ,\nonumber\\
X_{a}(t) &= \frac{1}{2}\left\{[L_{ax}(p_{1})-iL_{bx}(p_{1})]e^{p_{1}t}\right. \nonumber\\
&\left. \qquad   +[L_{ax}(p_{2})+iL_{bx}(p_{2})]e^{p_{2}t} \right\} ,\nonumber\\
Y_{a}(t) &= \frac{1}{2}\left\{[L_{ay}(p_{1})-iL_{by}(p_{1})]e^{p_{1}t}\right. \nonumber\\
&\left. \qquad   +[L_{ay}(p_{2})+iL_{by}(p_{2})]e^{p_{2}t} \right\} ,\nonumber\\
Y_{b}(t) &= \frac{1}{2}\left\{[L_{by}(p_{1})+iL_{ay}(p_{1})]e^{p_{1}t}\right. \nonumber\\
&\left. \qquad  +[L_{by}(p_{2})-iL_{ay}(p_{2})]e^{p_{2}t} \right\} .\label{ax}
\end{align}

\subsection{Solution of the Heisenberg-Langevin equations for the nonlinear coupling}\label{seca2}

The Laplace transform of the set of coupled differential equations for the quadratures components $X_{i}$ and $Y_{i}$ under the nonlinear coupling process, Eq.~(\ref{e35}), leads to the following sets of equations 
\begin{align}
\left(
\begin{array}{cc}
p+\kappa & -g  \\
-g & p+\kappa
\end{array}
\right) \left(
\begin{array}{c}
X_{b}(p)\\
X_{a}(p)
\end{array}
\right) = \left(
\begin{array}{c}
L_{bx}(p)\\
L_{ax}(p)
\end{array}
\right) 
\end{align}
and
\begin{align}
\left(
\begin{array}{cc}
p+\kappa & g  \\
g & p+\kappa
\end{array}
\right) \left(
\begin{array}{c}
Y_{a}(p)\\
Y_{b}(p)
\end{array}
\right) = \left(
\begin{array}{c}
L_{ay}(p)\\
L_{by}(p)
\end{array}
\right) 
\end{align}
Solving for $X_{b}(p)$ and $X_{a}(p)$, we find
\begin{align}
X_{a}(p) &= \frac{(p+\kappa)L_{ax}(p) +gL_{bx}(p)}{(p-p_{1})(p-p_{2})} ,\nonumber\\
X_{b}(p) &= \frac{(p+\kappa)L_{bx}(p) +gL_{ax}(p)}{(p-p_{1})(p-p_{2})}  , \label{b4}
\end{align}
and for  $Y_{a}(p)$ and $Y_{b}(p)$:
\begin{align}
Y_{a}(p) &= \frac{(p+\kappa)L_{ay}(p) -gL_{by}(p)}{(p-p_{1})(p-p_{2})} ,\nonumber\\
Y_{b}(p) &= \frac{(p+\kappa)L_{by}(p) -gL_{ay}(p)}{(p-p_{1})(p-p_{2})}  .\label{b5}
\end{align}
where 
\begin{align}
p_{1}= -\kappa +g ,\qquad p_{2} = -\kappa -g .
\end{align}
Taking the inverse Laplace transforms, we find that the time evolutions of the quadratures are
\begin{align}
X_{a}(t) &= \frac{1}{2}\left\{[L_{ax}(p_{1})+L_{bx}(p_{1})]e^{p_{1}t}\right. \nonumber\\
&\left. \qquad   +[L_{ax}(p_{2})-L_{bx}(p_{2})]e^{p_{2}t} \right\} ,\nonumber\\
X_{b}(t) &= \frac{1}{2}\left\{[L_{bx}(p_{1})+L_{ax}(p_{1})]e^{p_{1}t}\right. \nonumber\\
&\left. \qquad  +[L_{bx}(p_{2})-L_{ax}(p_{2})]e^{p_{2}t} \right\} ,\nonumber\\
Y_{a}(t) &= \frac{1}{2}\left\{[L_{ay}(p_{1})-L_{by}(p_{1})]e^{p_{1}t}\right. \nonumber\\
&\left. \qquad   +[L_{ay}(p_{2})+L_{by}(p_{2})]e^{p_{2}t} \right\} ,\nonumber\\
Y_{b}(t) &= \frac{1}{2}\left\{[L_{by}(p_{1})-L_{ay}(p_{1})]e^{p_{1}t}\right. \nonumber\\
&\left. \qquad  +[L_{by}(p_{2})+L_{ay}(p_{2})]e^{p_{2}t} \right\} .\label{bx}
\end{align}

\subsection{Populations of the modes and the correlation functions}

Having solutions for the quadrature components it is straightforward to evaluate populations of the modes, single-mode and two-mode correlation functions from the following relations
\begin{align}
\langle a^{\dag}a\rangle  &= \frac{1}{2}\left(\left\langle X^{2}_{a}(t)\right\rangle + \left\langle Y^{2}_{a}(t)\right\rangle - 1\right) ,\nonumber\\
\langle b^{\dag}b\rangle &= \frac{1}{2}\left(\left\langle X^{2}_{b}(\phi)\right\rangle + \left\langle Y^{2}_{b}(\phi)\right\rangle - 1\right) .
\end{align}
\begin{align}
\langle aa\rangle  &= \frac{1}{2}\left(\left\langle X^{2}_{a}\right\rangle - \left\langle Y^{2}_{a}\right\rangle+1+2i\left\langle X_{a}Y_{a}\right\rangle\right) ,\nonumber\\
\langle bb\rangle  &= \frac{1}{2}\left(\left\langle X^{2}_{b}\right\rangle - \left\langle Y^{2}_{b}\right\rangle+1+2i\left\langle X_{b}Y_{b}\right\rangle\right) .
\end{align}
\begin{align}
\langle a^{\dag}b\rangle  &= \frac{1}{2}\left[\left\langle X_{a}X_{b}\right\rangle \!+\! \left\langle Y_{a}Y_{b}\right\rangle 
+i\left(\left\langle X_{a}Y_{b}\right\rangle - \left\langle Y_{a}X_{b}\right\rangle\right)\right] ,\nonumber\\
\langle ab\rangle &= \frac{1}{2}\left[\left\langle X_{a}X_{b}\right\rangle - \left\langle Y_{a}Y_{b}\right\rangle +i\left(\left\langle X_{a}Y_{b}\right\rangle + \left\langle Y_{a}X_{b}\right\rangle\right)\right]  
\end{align}

\bibliography{RefJ}

\end{document}